\documentclass[graybox]{svmult}

\usepackage{mathptmx}       % selects Times Roman as basic font
\usepackage{helvet}         % selects Helvetica as sans-serif font
\usepackage{courier}        % selects Courier as typewriter font
\usepackage{type1cm}        % activate if the above 3 fonts are
\usepackage{makeidx}         % allows index generation
\usepackage{graphicx}        % standard LaTeX graphics tool
\usepackage{multicol}        % used for the two-column index
\usepackage[bottom]{footmisc}% places footnotes at page bottom
\usepackage{amsmath}
\usepackage{amssymb}
\usepackage{comment}

\makeindex             % used for the subject index
\def\beq{\begin{equation}}
\def\eeq{\end{equation}}
\def\be{\begin{equation}}
\def\ee{\end{equation}}

\def\cG0{{\cal G}_0}

\def\spinup{\uparrow}
\def\spindown{\downarrow}

\def\s{\sigma}
\def\uc2{$U_{c2}$}
\def\uc1{$U_{c1}$}

\def\bea{\begin{eqnarray}}
\def\eea{\end{eqnarray}}
\def \bal{\begin{align}}
\def \eal{\end{align}} %%aliases don't work in these environments, see ams FAQ
\def\#{\!\!}
\def\@{\!\!\!\!}

\def\+{\dagger}

\def\up{\spinup}
\def\down{\spindown}

\begin{document}

\title*{Weak \emph{and} strong electronic correlations in Fe superconductors}
% Use \titlerunning{Short Title} for an abbreviated version of
% your contribution title if the original one is too long
\author{Luca de' Medici}
% Use \authorrunning{Short Title} for an abbreviated version of
% your contribution title if the original one is too long
\institute{Luca de' Medici \at European Synchrotron Radiation Facility, BP 220, F-38043 Grenoble Cedex 9, France, \email{demedici@esrf.fr} \and Laboratoire de Physique et Etude des Mat\'eriaux, UMR8213 CNRS/ESPCI/UPMC, 10, rue Vauquelin, 75231 Paris, France}
%\institute{European Synchrotron Radiation Facility, BP 220, F-38043 Grenoble Cedex 9, France}
%%%%%%%%%%%%%%%%%%%%%%%%%%%%%%%%%%%%%
%TIIIIIIIIIIIIIIIIIIIIIIIIIIIIIIIIIIIIIIIIIPSSSSSSSSSSSS
%%%%%%%%%%%%%%%%%%%%%%%%%%%%%%%%%%%%%
% Use the package "url.sty" to avoid
% problems with special characters
% used in your e-mail or web address
%
% use \sectionmark{}
% to alter or adjust the section heading in the running head
%
% use \index{} to add index entries
%
%\runinhead{}  
%\subruninhead{}
% 
%\begin{svgraybox} \end{svgraybox}  %per le box grigie!!!

\maketitle

\abstract{  \newline
\indent
In this chapter the strength of electronic correlations in the normal phase of Fe-superconductors is discussed.
It will be shown that the agreement between a wealth of experiments and DFT+DMFT or similar approaches supports a scenario in which strongly-correlated and weakly-correlated electrons coexist in the conduction bands of these materials.
I will then reverse-engineer the realistic calculations and justify this scenario in terms of simpler behaviors easily interpreted through model results.
All pieces come together to show that Hund's coupling, besides being responsible for the electronic correlations even in absence of a strong Coulomb repulsion is also the origin of a subtle emergent behavior: orbital decoupling.
Indeed Hund's exchange decouples the charge excitations in the different Iron orbitals involved in the conduction bands thus causing an independent tuning of the degree of electronic correlation in each one of them. The latter becomes sensitive almost only to the offset of the orbital population from half-filling, where a Mott insulating state is invariably realized at these interaction strengths. Depending on the difference in orbital population a different 'Mottness' affects each orbital, and thus reflects in the conduction bands and in the Fermi surfaces depending on the orbital content.
}

\section{Introduction: electronic correlations?}
\label{sec:Intro}

Soon after the discovery of high-temperature superconductivity in the first iron pnictides a debate has sparked off, which is still lively to date: are the conduction electrons in these materials weakly \emph{or} strongly correlated?

This question is of fundamental importance on different levels. 

One is methodological, and it concerns finding the best theoretical viewpoint to model and predict the electronic properties of these compounds. Indeed mirroring the two sides of the conundrum two main points of view have polarized the community working on the subject. On one side the encouraging results of the standard one-body schemes such as Density Functional Theory (DFT), well capturing the topology of the Fermi surfaces and the main features of the bandstructure, have pushed many scientists towards weak-coupling approaches. Very generally, in these approaches the magnetism - experimentally found in the great majority of the stoichiometric compounds - is viewed as due to the nesting features of the Fermi surfaces and the superconductive pairing stems out of long wavelength collective excitations due to the proximity to the ordered state, such as spin-fluctuations.  
On the other side by postulating the vicinity of a Mott insulator and thus strong short-ranged electronic correlations, the magnetic phases of these materials were successfully modeled with a frustrated multi-orbital Heisenberg model (named "J1-J2"). Metallicity and spin-fluctuation mediated superconductivity in this view are obtained upon doping, in analogy with the single-band t-J model for cuprate superconductors.

Indeed analogies and parallels with these materials, which are the leading high-Tc superconductors, have been traced or sketched. The cuprates are the very paradigm for strongly correlated physics in the context of superconductivity, due to their phase diagram revolving around a Mott (charge-transfer) insulating antiferromagnetically ordered phase, and exotic bad-metallic properties defying an understanding as of today. 
The temptation to carry over all the wealth of new concepts and techniques developed in the 25-years attempt to understand the physics of these materials is natural, however to what extent this is possible depends obviously on the actual degree of correlation of Iron-based superconductors (Fe-SC).

Another important reason for which knowing the actual correlation strength is important, even having picked a side in the original debate, is the need of a quantitative theory. Indeed the two aforementioned starting points are both somewhat extreme. For instance the accord between observed Fermi surfaces and the ones calculated in the weak-coupling picture is substantially improved by including dynamical correlations (e.g. by Dynamical Mean-Field Theory - DMFT) that shift and renormalize the bands. This  obviously also affects the superconductive pairing, since the presence or absence of nodes is very sensitive to the detailed shape and size of the Fermi surface sheets, among other factors. Also some difficulties arise in weak-coupling approaches in reproducing the magnetic order in the iron chalcogenides.
On the other hand a quantitative estimate of the local magnetic moments postulated in the strong coupling approach is another aspect where the actual correlation strength matters and in this case too DMFT is a key tool for a realistic approach.

Indeed a possible settling of the initial viewpoint dichotomy is towards \emph{intermediate} correlations, which would not invalidate the weak-coupling band-structure and possibly bring it closer to the observed one \cite{Yin_kinetic_frustration_allFeSC}, while favoring the local quickly fluctuating moments bringing to the correct magnetism\cite{Hansmann_localmoment_prl}.
 
In this view an important role is played by the multi-orbital nature of  Fe-SC bands, and in particular by Hund's coupling, the intra-atomic exchange energy favoring the distribution of electrons in the same atomic shell on different orbitals with their spins aligned \cite{Haule_pnictides_NJP, Johannes_local-moment} which is known to be sizable in atomic Fe ($\sim 0.8 eV$).  Electronic correlations are indeed triggered by electron-electron interactions but in a multi-orbital context they are in a highly nontrivial relation to the Coulomb repulsion and Hund's coupling strengths.   
This means that the local Coulomb interaction energy U (of which Hund's J is a fraction) is likely not larger than the bandwidth, but the complexity introduced by Hund's coupling prevents this fact from implying weak correlations. In the "intermediate correlation" picture the role of Hund's coupling is mainly enhancing the effect of the local coulomb repulsion on the correlation strength. Besides, it is also of utter importance in determining the magnetic ground states\cite{Yin_Weiguo-Spin_fermion,Yin_kinetic_frustration_allFeSC}.

From the experimental point of view contrasting evidences have come, that can support weak, strong and intermediate correlation views. A few examples are:  hardly any Hubbard band has been detected by X-ray spectroscopies pointing towards weak correlations \cite{Yang_Devereaux_weakcorr}; the room-temperature resistivity is of the order the $m\Omega cm$, typical of strongly correlated bad metals \cite{Rullier_Hall_Resistivity}; reduction of the low energy optical spectral weight is moderate and has been taken as an evidence of intermediate correlations \cite{Qazilbash_correlations_pnictides}. 

The present chapter is devoted to the illustration of another possible solution of the puzzle, which is the presence of both strongly \emph{and} weakly correlated electrons, coexisting in the conduction bands. The various types of electrons showing up differently in various physical properties, this may explain many of the contrasting evidences on correlations, in a nutshell.

This view, originally put forth by the author and collaborators in Ref. \cite{demedici_3bandOSMT,demedici_Genesis} and postulated on a phenomenological basis in Ref.\cite{Kou_OSMT_pnictides,Hackl_Vojta_OSMT_pnictides,Yin_Weiguo-Spin_fermion,Yin_spinfermion_KFeSe} is more and more supported by realistic calculations using different techniques for treating the electronic correlations on top of DFT bandstructures: DMFT \cite{Haule_pnictides_NJP,Shorikov_LaFeAsO_OSMT,Laad_SusceptibilityPnictides_OSM,Craco_FeSe,Aichhorn_FeSe,Yin_kinetic_frustration_allFeSC,Ishida_Mott_d5_nFL_Fe-SC, Liebsch_FeSe_spinfreezing,Werner_122_dynU,Yin_PowerLaw}, variational Montecarlo\cite{Misawa_d5-proximity_magnetic}, Slave-spins mean-field (SSMF)\cite{YuSi_LDA-SlaveSpins_LaFeAsO,Yu_Si_KFeSe}, Hartree-Fock mean-field \cite{Bascones_OSMT_Gap_halffilling}, fluctuation-exchange approximation \cite{Ikeda_pnictides_FLEX}, Gutzwiller approximation \cite{Lanata_FeSe_LDA+Gutz}.

However it is clear that, to date, the predictivity of full \emph{ab initio} approaches for correlated materials is still under development (particularly due to the presence of "double counting" corrections for the electronic interactions, see section \ref{elstruct}). Thus the approach chosen here is rather to use theoretical guidance by these realistic approaches to harvest the wealth of available experimental data in search of the main trends as far as electronic correlations are concerned.

\begin{figure}[t]
\sidecaption
\includegraphics[width=7.cm]{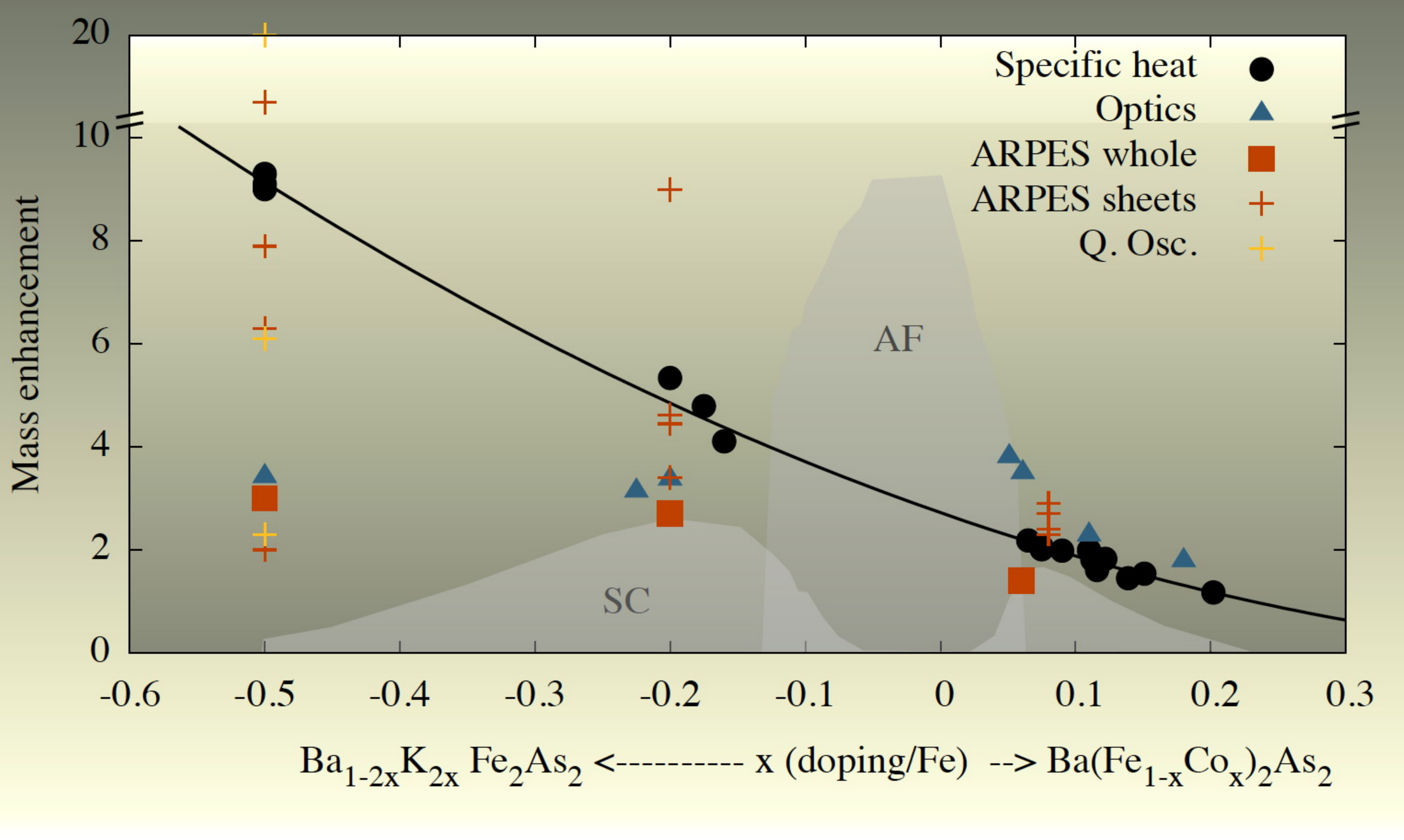}
\caption{Experimental mass enhancement estimates in K- and Co- doped BaFe$_2$As$_2$ from different techniques (see legend). Different ARPES and Quantum Oscillation data at the same doping represent the estimate for the various sheets of the Fermi surface. The spread of these estimates increases with hole-doping and can be interpreted as an increasing selective Mottness (see text). From \cite{demedici_OSM_FeSC}.}
\label{fig:Exp_OSM_FeSC}
\end{figure}

\begin{figure}[t]
\sidecaption
\includegraphics[width=7.cm]{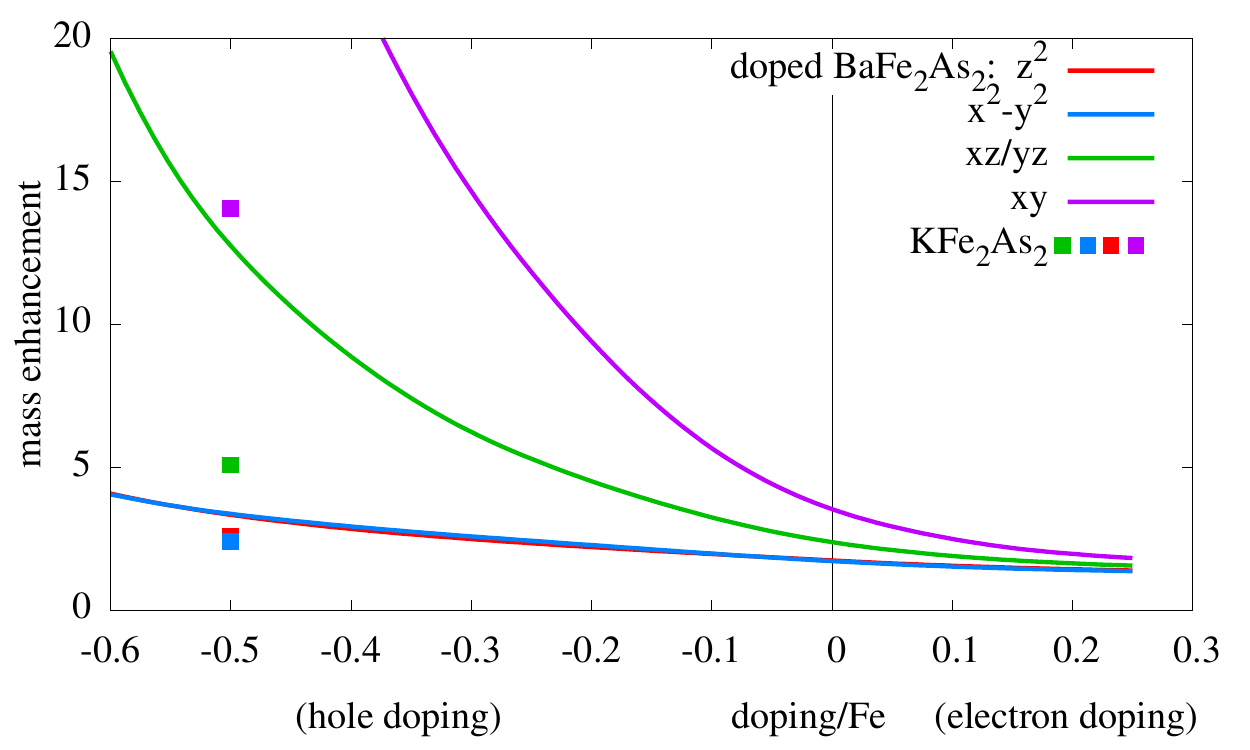}
\caption{Theoretical mass enhancements of the conduction electrons by Fe-orbital character, obtained in DFT+Slave-spin mean-field (SSMF\cite{demedici_Slave-spins,Hassan_CSSMF}) for doped BaFe$_2$As$_2$ (lines) and at filling of 5.5 electrons for the structure of KFe$_2$As$_2$ (squares). The increase and spreading of the orbitally-resolved correlation strengths parallels the experimental one reported in Fig. \ref{fig:Exp_OSM_FeSC}. From  \cite{demedici_OSM_FeSC}.}
\label{fig:Theo_OSM_FeSC}   
\end{figure}

A full analysis of experiments in parallel with theory (DFT+Slave-spin\cite{demedici_Slave-spins,Hassan_CSSMF}), supporting this scenario was performed on (both hole- and electron-) doped BaFe$_2$As$_2$ by the author together with G. Giovannetti and M. Capone in Ref. \cite{demedici_OSM_FeSC}, and is summarized in Figures \ref{fig:Exp_OSM_FeSC} and \ref{fig:Theo_OSM_FeSC}. The experimental estimates of the mass enhancements induced by correlations show two main trends: with reducing filling of the conduction bands they globally increase, and they spread. This trend culminates for the stoichiometric end-member KFe${}_2$As$_2$ (half a hole of doping per Fe ion compared to BaFe${}_2$As$_2$) in which the masses of the carriers are so different that a heavy-fermionic behaviour is realized\cite{Hardy_KFe2As2_Heavy_Fermion}.
This is exactly what the theoretical calculations predict (see Fig. \ref{fig:Theo_OSM_FeSC}) and the reason for this behaviour will be clarified in detail in this chapter. 

%%%%%%%%%%%%%%%%%%%%%%%%%%%%%%%%%%%%%%%%%%
%%%%%%%%%%%%%%%%%%%%%%%%%%%%%%%%%%%%%%%%%%

Indeed in the same work Ref. \cite{demedici_OSM_FeSC} it was pointed out that the general mechanism behind this is the "orbital decoupling"\cite{demedici_MottHund} induced by Hund's coupling. This is a general \emph{emergent phenomenon} in which the charge excitations in different orbitals become decoupled under certain conditions (the most important being a sizable Hund's coupling) and correlations are tuned in every orbital in an essentially independent way. The main variable for tuning the correlation in each orbital becomes the doping of that orbital by respect to individual half-filling. Then the \emph{final} orbital populations (i.e. those in the interacting system, in contrast to the bare orbital populations in absence of dynamical correlations) determine the correlation strength in each orbital (this has been termed \emph{selective Mottness}).  

This robust behavior is found common to all Fe-SC investigated thus far and an essential tool to understand the final correlation strengths in terms of intermediate variables.
In this sense it is an emerging phenomenon: when this regime is realized, whatever the initial microscopic parameters, the final correlation strengths will almost solely be determined by the respective orbital populations, which are quasi-independent emergent variables.
This correlation between variables on a hierarchical level just above the purely microscopic one, may have an advantage in an attempt of designing new materials with interesting emergent properties over, i.e. tuning single hopping elements or distances and angles in the atomic structure.

In the following I will use previous and new results on model systems to "reverse-engineer" this result found in Fe-SC, break it down in simpler well understood results, and show that the orbital decoupling is indeed the mechanism at play.
%%%%%%%%%%%%%%%%%%%%%%%%%%%%%%%%%%%%%%%%%%
%%%%%%%%%%%%%%%%%%%%%%%%%%%%%%%%%%%%%%%%%

This has a double utility. 
On one hand it validates the generality and robustness of the phenomenon - that can indeed be found in a one-shot ab-initio calculation - to the variation of microscopic parameters, such as the screened Coulomb interaction strengths, whose estimates are still subject to a sizable uncertainty.
On the other hand it paves the way towards the conception of low-energy models, retaining all and only the essential elements in order e.g. to model the superconductive pairing and the magnetic order.

Finally I will briefly mention another result of Ref. \cite{demedici_OSM_FeSC} that is the analogy between Fe-SC and high-Tc cuprates that can be traced based on the orbital-decoupling mechanism. Indeed it can be shown within a certain theoretical approach to the Hubbard model (the Dynamical cluster approximation, one of the cluster extensions of DMFT), that correlations differentiate \emph{in k-space} following the same dependence on the population of that part of the Brillouin zone, analogously to the multi-orbital case. 
This suggests an intriguing and deep parallel between the phase diagrams of the two kind of superconductors and the possible importance of selective Mottness for high-Tc superconductivity.

The plan of the chapter is the following. In Section \ref{elstruct} I will recall the basic features governing the low-energy physics in FeSC, from the electronic structure to the treatment and intensity of electronic correlations. In Section \ref{sec:overall_corr} I will discuss the overall degree of correlation of the conduction electrons, outlining the central role of atomic Hund's coupling, highlighting the usefulness of the simple yet very appealing Hubbard criterion for the occurrence of the Mott transitions. I will recall the large body of realistic calculations which include electron-electron correlations explicitly and point out their main features, including the general tendency towards orbital differentiation of the correlation strength. In Section \ref{sec:exp_abinitio_OSM} the experimental evidences supporting the coexistence of more strongly and more weakly correlated electrons in FeSC are gathered and analyzed in some detail, in the light of realistic calculations and discussion of Ref. \cite{demedici_OSM_FeSC}. In Section \ref{sec:orb_dec} the mechanism behind the differentiation of correlations, the Hund's induced orbital decoupling, is reverse-engineered in terms of the basic behavior found in simpler models, and in the light of a generalized Hubbard criterion representing a stylization of the physics of Hund's metals in the regime near the Mott insulator realized at half-filling. In Section \ref{sec:realistic_Orb_dec} I go back to the realistic calculations illustrating how this cartoon applies to the case of FeSC and use two 'wrong' realistic calculations to point out the role of the microscopic features of the bandstructure on the final result.
Finally in Section \ref{sec:conclusions} conclusions are drawn, and some of the issues left out of this chapter are briefly mentioned. Some further calculations analyzing the details of the orbital-decoupling mechanism in models are reported in the appendix.

\section{Essentials of the electronic structure of Fe-based pnictides and chalcogenides}\label{elstruct}

Iron-based superconductors are formed by stacked Fe-pnictogen or Fe-chalcogen planes, with (for the former) or without (for the latter) filler planes providing charges. The Fe atoms form a square lattice with the ligand positioned in the middle of each square, alternatively above or below the plane. The formed structure has tetragonal symmetry, which undergoes an orthorhombic distortion in proximity of the magnetic phase. 

Throughout the chapter we will focus on the normal tetragonal phase, that is where superconductivity is realized, at low temperatures.
In this structure the bands cutting the Fermi level are mainly of Iron 3d orbital character with a sizable admixture of the p-orbitals from the ligands. The bandwidth of this complex is around $4-5 eV$, while the bonding bands, with predominant character of the ligand p-orbitals, lie just below and extend for $3-4 eV$.
The Fermi surface reflects the semi-compensated metallic band structure of the low-energy complex, and hole pockets are formed around the Brillouin zone center while electron pockets around the zone corner\footnote{Two conventions are typically used for the unit cell and the consequent Brillouin zone. Depending on the convention used the electron pockets are centere either on the corner (e.g. $(\pi,\pi)$), or on the side  (e.g. $(\pi,0)$) of the Brillouin zone. The present discussion is independent of the convention used.}.

Albeit the DFT (in its most common Local Density Approximation - LDA, or Generalized Gradient Approximation, GGA) bandstructure includes a mean-field electron-electron interaction effect on the one-electron wave functions, in order to incorporate the dynamical many-body correlations one way is to construct a Hubbard-like low-energy model, in which local (multi-orbital) interactions are explicitly treated. In order to do this one has to construct a local basis (one typical choice is using maximally localized Wannier orbitals) and re-express the bandstructure through a tight-binding model, on top of which one adds the Hubbard local interaction terms. 
The static contribution of the electron-electron interaction is thus counted twice and a double counting (interaction-dependent) energy has to be subtracted. 

Here, two choices are possible, customarily. One is to use a larger basis of local orbitals, including explicitly both the correlated (i.e. on which the Hubbard term will be acting) d-orbitals of main Fe character, and the non-interacting p-orbitals of main ligand character, so that the tight-binding bandstructure reproduces the bands over the whole ("large") energy window of $\sim8-10eV$ around the Fermi level $E_F$. In this formulation the local orbitals are very atomic-like, thus better justifying the d-orbital Hubbard interaction terms, but the double-counting term, acting only on the correlated orbitals, alters the energy distance between the d and the p orbitals. Drawbacks due this last issue have been highlighted lately, and different prescriptions in order to fix it have been proposed. Then, albeit this formulation is in principle more realistic, the possibility of a completely ab-initio approach is still a matter of  ongoing research\cite{Wang_covalency,Parragh_d-dp_models,Hansmann_Udp,Dang_doublecounting,Haule_all_electron_DMFT,Hansmann_Upd_cuprates}.

In order to avoid this problem we will only discuss the alternative formulation that uses a smaller basis of local orbitals, including only correlated d-symmetry orbitals of main Fe character, so that the tight-binding bandstructures reproduces the low-energy complex over a ("small") 4-5eV  window around $E_F$. These local orbitals are more extended in space compared to the ones in the large-window formulation but the double-counting term is simply absorbed in the chemical potential that is fixed by the choice of the conduction electron population. Thus the double counting problem is altogether avoided. We will see that the reliability of this (a priori less accurate) choice is justified in our case by the agreement with the experimental data.

The resulting 5-orbital model is populated by 6 electrons in the stoichiometric Fe-SC and its tight-binding parameters have the main following features.
The cubic environment splits the orbital energies in a lower doublet (of $e_g$ symmetry) and an upper triplet (of $t_{2g}$ symmetry). These are further split by the tetragonal symmetry so that only the two $t_{2g}$ orbitals with lobes pointing out of plane (called xz and yz) remain degenerate. Overall, however, these crystal-field splittings are one order of magnitude smaller than the bandwidth, the orbital energies being spread over an interval of $0.3-0.5 eV$. This implies that all 5 orbitals will participate to the conduction bands and to the Fermi surfaces, even if it is found that these are predominantly of $t_{2g}$ character.
%%%%%%%%%%%%%%%%%%%%%%%%%%%%%%%%%%%%%%%%%%
\begin{figure}[h]
%\sidecaption
\begin{center}
\includegraphics[width=11.cm]{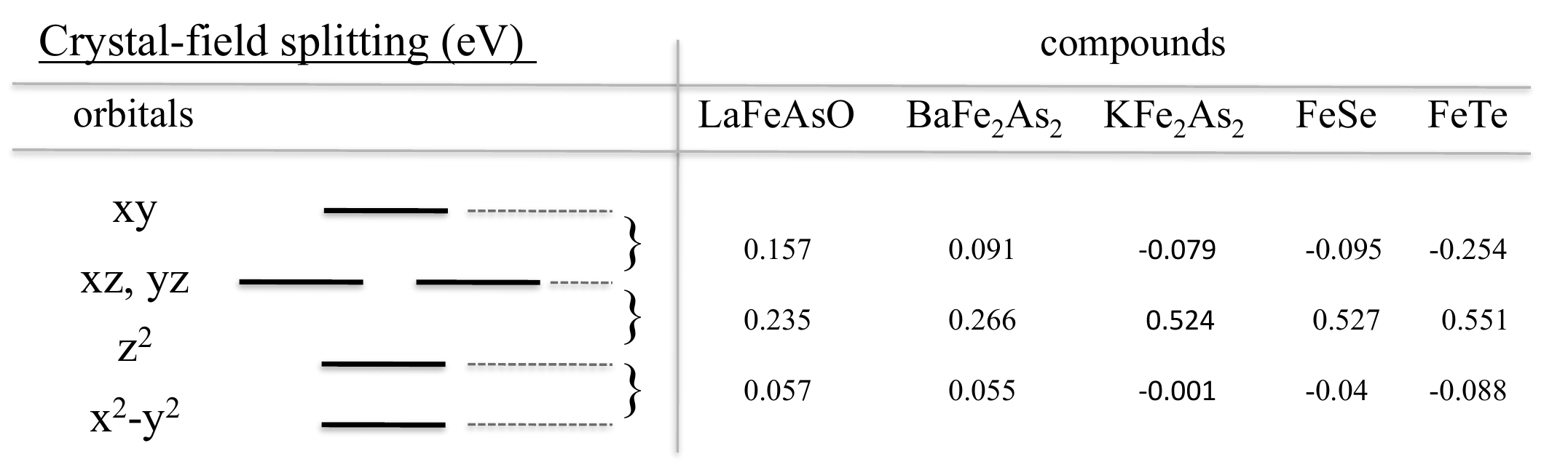}
\caption{Crystal field splitting of the orbital levels in the tight binding parametrization of various stoichiometric Fe-SC (technical details can be found in \cite{demedici_OSM_FeSC}).}
\end{center}
\label{fig:Crystal_fields}   
\end{figure}
%%%%%%%%%%%%%%%%%%%%%%%%%%%%%%%%%%%%%%%%%
As can be seen in the table in Fig. \ref{fig:Crystal_fields} the most important splitting is between the $t_{2g}$ and $e_{g}$ multiplets, the intra-multiplet splittings are smaller and can change sign across the compounds.

%%%%%%%%%%%%%%%%%%%%%%%%%%%%%%%%%%%%%%%%%%
\begin{figure}[h]
\sidecaption
\includegraphics[width=7.5cm]{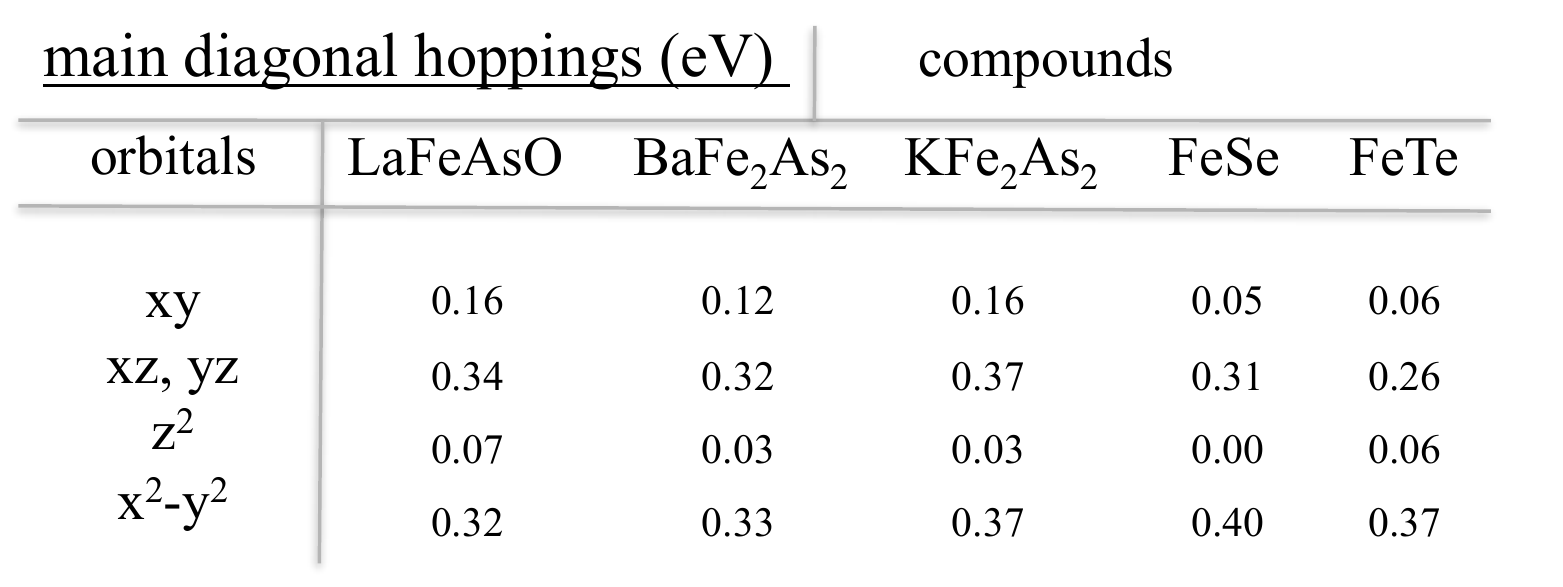}
\caption{As in the previous figure, but main diagonal in-plane nearest-neighbor hoppings. Notice that for xz and yz orbitals these are very directional, hence here the $t^{100}_{xz-xz}=t^{010}_{yz-yz}$ are reported.}
\label{fig:main_hops}   
\end{figure}
%%%%%%%%%%%%%%%%%%%%%%%%%%%%%%%%%%%%%%%%%

The table in Fig. \ref{fig:main_hops} shows instead the main diagonal (i.e. conserving the orbital nature) nearest-neighbor in-plane hoppings (next-nearest neighbor hoppings are of the same order but tipically somewhat smaller). The main features to be highlighted are that $x^2-y^2$ has larger in plane hoppings than $z^2$ and most importantly that $xz/yz$ have larger in-plane hoppings than $xy$. The smallness of the $xy-xy$ hopping has been related to the angle formed by the bonds between the Fe ion and the ligands in \cite{Yin_kinetic_frustration_allFeSC}. A cancellation (named "kinetic frustration") happens between the direct and indirect (i.e. through the ligand) amplitudes, reaching very small values in the chalcogenides where the height of the ligand is maximal.

Inter-orbital hoppings are quite large, and connect all the 5 orbitals together forming the intricate final bandstructure.
Inter-plane hoppings are typically much smaller, there is however a non-negligible band dispersion in the z direction due mainly to the $z^2$ orbital.

The Hubbard hamiltonian $H=H_0+H_{int}$ used includes the tight-binding one-electron part reading:
\begin{equation}
\label{Ham}
H_0=\sum_{ij,ll^\prime,\sigma}t^{ll^\prime}_{ij}c^\dag_{il\sigma}c_{jl^\prime\sigma}-\sum_{i,l,\sigma}\mu c^\dag_{il\sigma}c_{il\sigma},
\end{equation}
where $c^{\dag}_{il\sigma}$ and $c_{il\sigma}$ are the fermionic creation and annihilation operators acting on site $i$, orbital $l$ 
and spin $\sigma$ and $t^{ll^\prime}_{ij}$ is the hopping amplitude between orbital $l$ on site $i$ and orbital $l^\prime$ on site $j$. 
$H_{int}$ describes the local multi-orbital electron-electron Coulomb interaction and customarily the rotationally invariant Kanamori Hamiltonian is used, that reads
 \bea\label{H_int}
 H_{int}\@&\,=\,\@&U \sum_{il}
 n_{il\up} n_{il\down}+(U-2J)\sum_{i,l>l^\prime,\s} n_{il\s}
  n_{il^\prime\bar\s} + (U-3J)\sum_{i,l>l^\prime,\s} n_{il\s}  n_{il^\prime\s}\nonumber \\
 \@&-\@&J\sum_{i,l\neq l^\prime}\left[d^\+_{il\up}d_{il\down}d^\+_{il^\prime\down}d_{il^\prime\up}+d^\+_{il\up}d^\+_{il\down}d_{il^\prime\up}d_{il^\prime\down}\right],
 \eea
where $n_{il\s}=c^\dag_{il\sigma}c_{il\sigma}$, $U$ is the strength of intra-orbital Coulomb repulsion, $U-2J$ that of the inter-orbital one, and J is the Hund's coupling. This form, which is exact separately for a $t_{2g}$ and a $e_{g}$ multiplet \cite{Georges_Annrev}, is a common approximation for a 5-orbital 3d shell. Moreover in a realistic framework the matrix elements are in general orbital-dependent. However no result in the Fe-SC framework has been reported in the author's knowledge, where these differences play a major role, and the discussion of these terms is beyond the scope of this chapter where a reductionist approach is adopted. Another commonly used approximation is dropping the last two terms in (\ref{H_int}) and keeping only the density density terms. 

A full discussion of the ab-initio estimates of U and J can be found in \cite{miyake_interactions_jpsj_2010}. 
%%%%%%%%%%%%%%%%%%%%%%%%%%%%%%%%%%%%%%%%%%
\begin{figure}[h]
\sidecaption
\includegraphics[width=6cm]{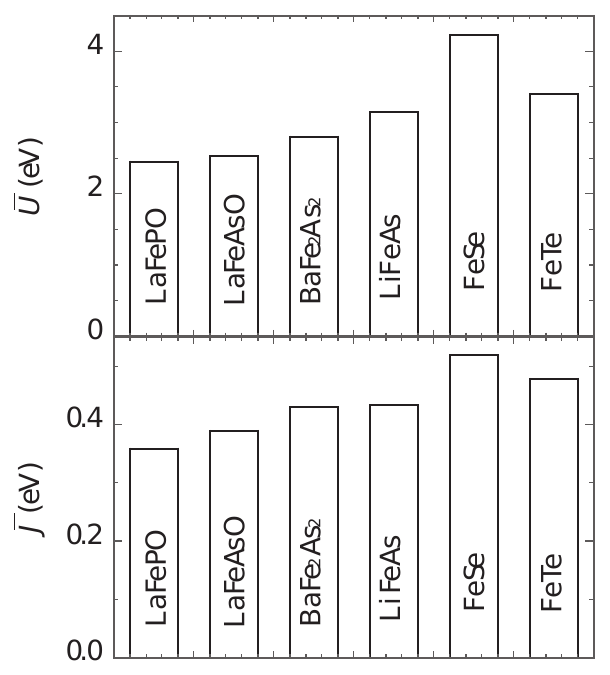}
\caption{Orbitally averaged ab-initio interaction parameters for various Fe-SC. Adapted from \cite{miyake_interactions_jpsj_2010}.}
\label{fig:Miyake}   
\end{figure}
%%%%%%%%%%%%%%%%%%%%%%%%%%%%%%%%%%%%%%%%%
We report the orbitally-averaged values of U and J estimated in that work by T. Miyake et al. in Fig. \ref{fig:Miyake}. It can be noted that $U\sim 2.5 - 4 eV$ and $J\sim0.35 - 0.5 eV$, with the maximum values reached in the chalcogenides.

\section{Overall correlation strength: the "Janus" effect of Hund's coupling}\label{sec:overall_corr}

By looking at the energies at play in Fe-SC low-energy physics as outlined in the previous section one may naively vouch for a weakly correlated regime\cite{Yang_Devereaux_weakcorr}. Indeed the Hubbard repulsion $U$ is smaller than the total bandwidth $W$, which is the usual zeroth-order rule of thumb to assess the correlation strength in a Hubbard-like system. This criterion complies with obvious energetic arguments (the kinetic energy, roughy measured by the bandwidth is opposed by the interaction energy, measured by U) and most concretely is tailored around the simplest estimate of the critical interaction strength $U_c$ needed to have a Mott insulating state in the half-filled single-band Hubbard model. 

\begin{figure}[h]
%\sidecaption
\begin{center}
\includegraphics[width=8cm]{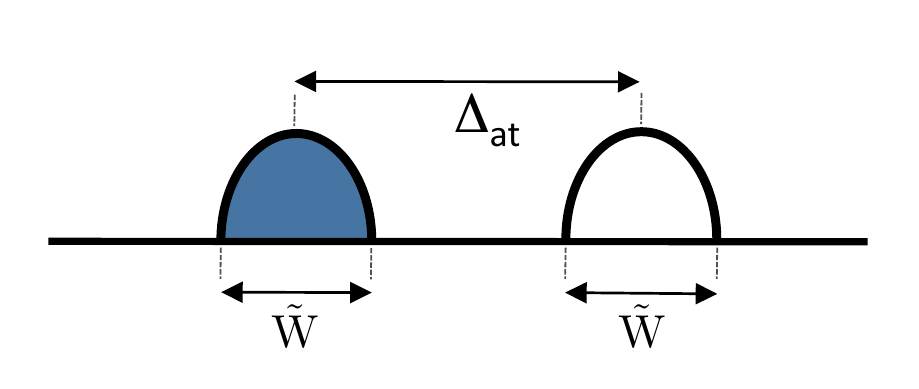}
\end{center}
\caption{Hubbard bands emanate from the atomic excitation spectrum. They are spread apart following the atomic Mott gap $\Delta_{at}$ (i.e. the sum of the energies needed to create a particle excitation and a hole excitation), and are broadened by an effective bandwidth $\tilde{W}$.  In the single-band Hubbard model $\Delta_{at}=U$ and $\tilde{W}=W$. Here depicted is the half-filled case in the Mott insulating phase $U>U_c$. The dark color represent "filled states" (the extraction part of the spectrum).}
\label{fig:Hubbard_bands}   
\end{figure}

This strong-coupling argument, due to Hubbard himself \cite{Hubbard_III} and illustrated in Fig. \ref{fig:Hubbard_bands}, is based on the fact that the excitation spectrum of a Mott insulator is roughly the atomic-limit spectrum broadened by the hopping amplitudes, and it describes the atomic charge excitations that can incoherently propagate through the system. In the single band model then it is mainly formed by two "Hubbard bands" at distance U (the atomic excitation energy) from one another (and symmetrically placed around the zero-energy point in the particle-hole symmetric half-filled case) that can be shown to disperse on an energy range $W$. Then an insulator-to-metal transition is obtained when U is reduced to a point that the two bands overlap, the gap closes and spectral weight is brought back to zero energy. This yields $U_c=W$ in the single-band case. At any $U>U_c$ a metallic state can also be restored by doping: in this simple picture this is equivalent to bringing the chemical potential, across the gap, into the Hubbard bands, where quasiparticle states are then created. In the proximity of these Mott transitions the metallic state is expected to be strongly correlated (mean-field studies show that the quasiparticle weight Z vanishes as $|U-U_c|$ below the transition and proportionally to carrier doping above $U_c$\cite{Brinkman_Rice,Coleman_Slave_bosons, kotliar_ruckenstein, Zitko_extremely_correlated_Fermi_Liquids})
%%%%%%%%%%%%%%%%%%%%%%%%%%%%%%%%%%%%%%%%%%
%%%%%%%%%%%%%%%%%%%%%%%%%%%%%%%%%%%%%%%%%

In a Hubbard model of M degenerate orbitals, where Mott transitions can happen at any integer filling, the effective kinetic energy is enhanced by orbital fluctuations (the Hubbard bands disperse on an energy range $\tilde{W}\sim \sqrt{M}W$), thus increasing the $U_c$ \cite{Lu_Gutz_multiorb, gunnarsson_fullerenes, Rozenberg_multiorb, Florens_multiorb}. At fixed interaction strength, correlations are reduced accordingly and this is the reason, for instance, believed to account for the metallic state of the 3-fold degenerate system $A_3C_{60}$ ($A$=K, Rb) in which $U\sim1.5 W$, thus substantially larger than the bare bandwidth \cite{gunnarsson_fullerenes}.

Thus applying naively the generalization to 5 orbitals of the Hubbard criterion (where $U_c=\tilde{W}\sim 2W$) one may conclude that Fe-SC, in which U is well below this estimate, are weakly correlated.
However one should refrain from assessing the correlations of the metallic state based solely on the comparison of the interaction strength U and the bandwidth W, even in a simplified treatment, for two reasons. First, because as we will see a third scale, the Hund's coupling J, plays a major role, not least in modifying the Hubbard criterion itself. Second because, albeit in the proximity of a Mott insulator one can expect a metallic state to be strongly correlated, outside this range the Hubbard criterion is not necessarily useful and one has to rely on more quantitative approaches.
A more reliable method in this respect is DMFT that can describe both the metallic and insulating phases on the same footing.

In the dynamical mean-field language\cite{georges_RMP_dmft} a site in the lattice system is described as a quantum impurity exchanging electrons with an effective bath that represents the rest of the system. The metallic phase corresponds to the Kondo-screening of the local moment  induced by the local interaction and the Mott transition is then described as a screening/unscreening transition. In the screened phase the impurity model (and hence the lattice system) has a local Fermi-liquid self-energy, and the degree of coherence of the corresponding quasiparticles (and hence the mass of the correlated electrons) in the metal is characterized by the Kondo temperature $T_K\propto\exp(-1/[2 \rho J_K])$, where $\rho$ is the conduction-electron density of states per spin at the Fermi energy and $J_K$ is the antiferromagnetic Kondo coupling. This effective coupling is due to the exchange energy gained in the processes of resonant scattering of the conduction electrons on the impurity\footnote{For the Mott transition, an intuitive connection between the screening/unscreening process and the Hubbard criterion is made by the self-consistent nature of the effective DMFT bath. When the low-energy coherence is too low, it is convenient for the system to lower the energy of the low-lying filled states by opening a gap at the Fermi energy ($\rho=0$) and form a Mott insulator, which is self-consistent because when $\rho=0$, $T_K=0$ (for a discussion see e.g. \cite{Nozieres_comments_Mott_Transition})}.
  
The aforementioned increase in kinetic energy for a degenerate system of M orbitals corresponds here to an increase of the effective Kondo coupling $J^M_{K,eff}=MJ^1_K$, since the impurity can exchange with the bath both spin and angular momentum, and this facilitates the screening. 
The consequent increase of the Kondo temperature is reflected in an increased overall coherence of the conduction electrons\cite{hewson_book_1993,coqblin_schrieffer_1969,okada73}.

However Hund's coupling strongly modifies this picture. Indeed the effect of this coupling is to favor, among the many atomic degenerate states with a given total charge, those with a larger total spin S, and among these the ones with larger total angular momentum L.
This considerably lowers the degeneracy of the ground state, and thus the channels for exchange processes with the bath, and ultimately the effective Kondo coupling. 

For instance at large $J$ for a half-filled shell the effect is maximal and $J^M_{K,eff}=J^1_K/M$ \cite{okada73}.
The reduction of $M^2$ in this case compared to the enhanced case without Hund's coupling is understood intuitively, due to the complete quenching of the angular momentum (the multiplet of maximum S is bound to have $L=0$ in a half-filled shell): the factor M of enhancement compared to the single-orbital case is lost since orbital exchange is blocked, and another factor 1/M is due to the reduced spin exchange due to the selection of the high-spin multiplet for a ground state\cite{schrieffer_japplphys_1967,nevidomskiy_coleman_Kondo_Hund}. 

 For a shell with $M\pm1$ filling (which is the case of stoichiometric Fe-SC) spin and orbital exchange channels at fixed valence have to be considered, ultimately still leading to a strong reduction of the $T_K$, similarly to the half-filled case \cite{okada73,Yin_PowerLaw,Aron_RenGroup_Hunds_metals}. Valence fluctuations could add even further complexity to this analysis. However it is clear that for shells near or at half-filling Hund's coupling has a strong effect in reducing the coherence scale of conduction electrons in the metallic phase (a more thorough discussion of these topics can be found in Ref. \cite{Georges_Annrev}).

This strengthening of correlations could in principle simply favor a Mott insulating state, however this is not true in general. 
In fact besides this low-energy effect Hund's coupling has another influence on the system, stemming directly from the high-energy atomic features of the spectrum. Indeed Hund's rules are first and foremost an atomic effect, and the selection of the low-lying multiplets in each sector with a given total atomic charge has also an effect on the atomic Mott gap.
The gain in energy due to J in every charge sector is different, thus the corresponding Mott gap is modified differently. In practice the half-filled sector is the one that has the maximal gain in energy, and hence the Mott gap for a half-filled shell is $\Delta_{at}=U+(M-1)J$ and is always enhanced by J. For all other fillings however the gap value is $\Delta_{at}=U-3J$ and is actually \emph{reduced} by Hund's coupling\cite{vandermarel_sawatzky_prb_1988,vandermarel_phd_1985}.

Given the aforementioned relation between the atomic spectrum and the one of a Mott insulator, one can again apply a Hubbard-like criterion (Fig. \ref{fig:Hubbard_bands}) and see that in the half-filled case the opening of a Mott gap will be favored by J (and $U_c$ will be reduced), while in all other cases the Mott insulating state will be pushed away at very high $U_c$ \cite{demedici_MottHund}\footnote{Indeed it is found numerically that the values of $U_c$ at large J scale well with $\Delta_{at}$ \cite{demedici_MottHund}. It can be shown (at least in specific cases) that at large J the effective width of the Hubbard bands $\tilde{W}$ tends to a constant (the single-band value $\tilde{W}\simeq W$ for the half-filled case - de' Medici and Capone, unpublished) owing to the quenching of orbital fluctuations. At small J instead orbital fluctuations are still active and their reduction with J (and the consequent reduction of $\tilde{W}(J)$) dominates over the tuning of $\Delta_{at}(J)$. Thus while in the half-filled case the two effects add up, causing an even faster reduction of $U_c$, in the "Janus" case (see the main text) they work against one-another (not surprisingly, since the reduction of $\tilde{W}$ is related to the loss of kinetic energy due to the reduction of the $T_K$) causing an initial decrease of $U_c$ before a strong increase.}.
%%%%%%%%%%%%%%%%%%%%%%%%%%%%%%%%%%%%%%%%%%
\begin{figure}[h]
\sidecaption
\includegraphics[width=11.7cm]{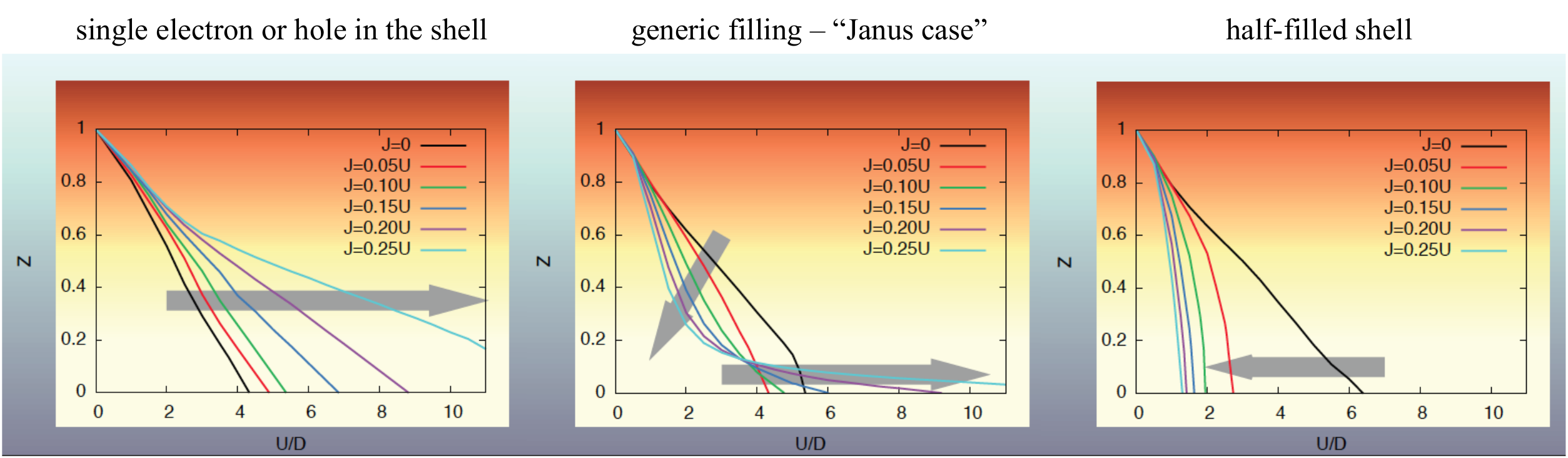}
\caption{Influence of Hund's coupling (grey arrows) on the quasiparticle weight in a (here - particle-hole symmetric, 3-orbital) degenerate Hubbard model. Three cases are possible depending on the filling. At half-filling (right panel) the Mott insulating state is strongly favored, for a filling of a single electron or hole (left panel) the metallic state is favored. For all other fillings (middle panel) coherence is reduced but the Mott insulator is disfavored, stabilizing a strongly-correlated (typically bad-metallic) phase for a large range of parameters. Adapted from Ref. \cite{demedici_Janus} (these results hold analogously for a 5-band Hubbard model \cite{Lanata_FeSe_LDA+Gutz}).}
\label{fig:Janus}   
\end{figure}
%%%%%%%%%%%%%%%%%%%%%%%%%%%%%%%%%%%%%%%%%
  
For half-filled systems this effect collaborates with the reduced Kondo screening in enhancing correlations and favoring strongly the Mott insulating state\cite{pruschke_Hund} (right panel in Fig. \ref{fig:Janus}). However for all other filling the two effects are antagonistic to one another\footnote{In the limiting case of a shell populated by only one electron or one hole per site however, the low-energy effect of Hund's coupling is absent, so that the correlation strength is simply reduced following the enhancement of $U_c$ (left panel in Fig. \ref{fig:Janus})}.
This case in which Hund's coupling has two antagonistic effects was nicknamed after the double-faced god "Janus" from the roman mythology in Ref. \cite{demedici_Janus}.

In the "Janus" case, when J/U is sizable, correlations are quickly enhanced by interactions and the system acquires a low coherence temperature and large mass enhancement already at quite low values of U. However the Mott insulating state is pushed away, since Hund's coupling keeps the Hubbard bands from spreading apart and thus forces spectral weight to low energy, even if with reduced coherence. This state is realized for a large range of interaction strength and the quasiparticle weight $Z(U)$ shows the typical S-shaped form of the middle panel of Fig. \ref{fig:Janus}.

%%%%%%%%%%%%%%%%%%%%%%%%%%%%%%%%%%%%%%%%%%
\begin{figure}[h]
\sidecaption
\includegraphics[width=11.7cm]{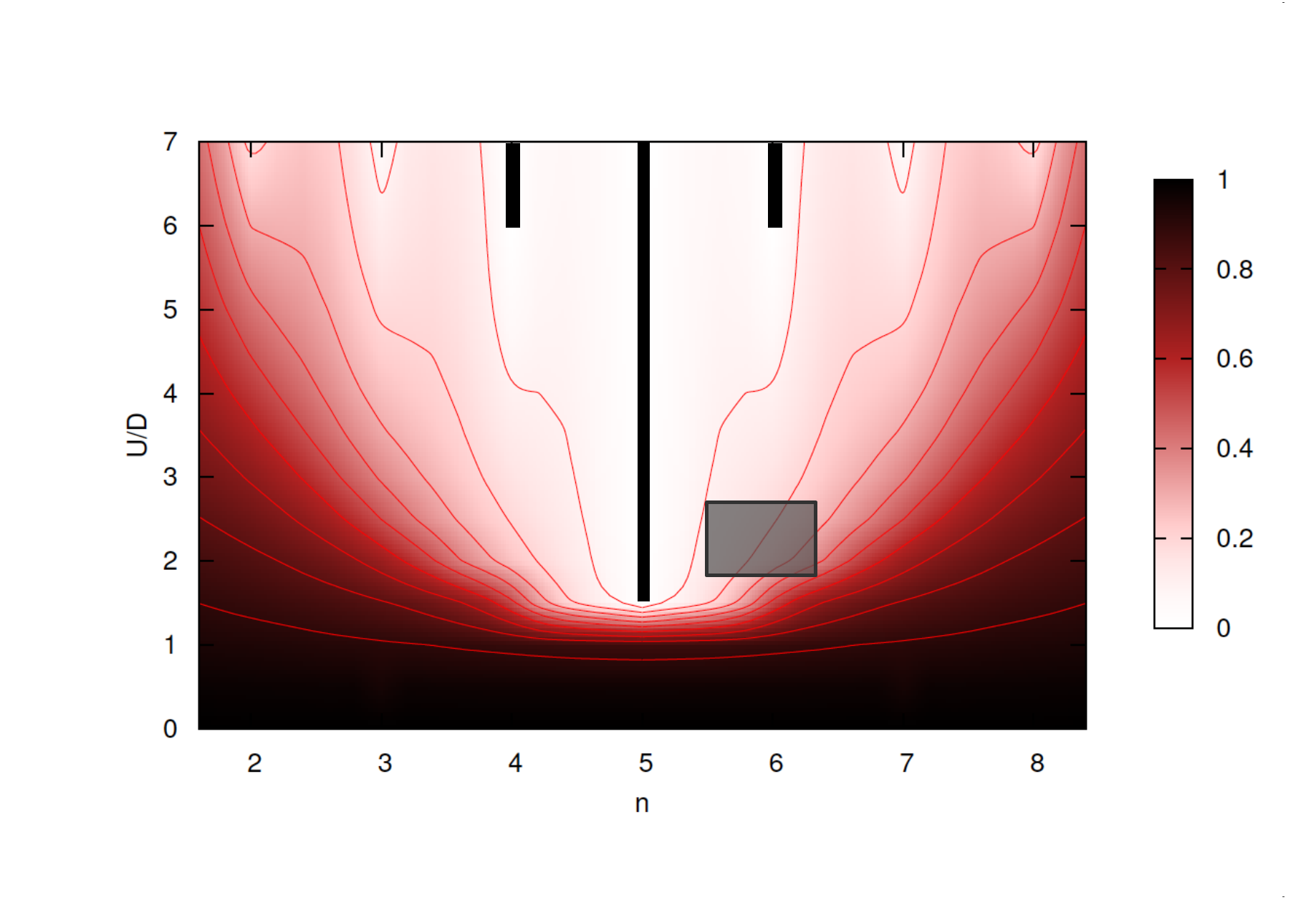}
\caption{Quasiparticle weight (color code) as a function of filling and interaction strength U at fixed J/U=0.2 (see footnote \ref{foot:ratioJU} for the realistic $J/U$ value) for a 5-band degenerate Hubbard model (the bands have semi-circular densities of states of half-bandwidth D) solved within slave-spin mean-field. Even if this model does not take into account the specificities of the band-structures of the Fe-SC, it gives nevertheless an idea of the proximity of the Mott transitions (signaled by the black bars) to the realistic range of parameters for Fe-SC (gray square, values of U are normalized to the bare Kinetic energy of the bandstructures in order to be compared to this model calculation).}
\label{fig:5bands}   
\end{figure}
%%%%%%%%%%%%%%%%%%%%%%%%%%%%%%%%%%%%%%%%%
It can also be shown that the Janus effect is more and more pronounced with increasing number of orbitals $M\ge 3$\cite{demedici_Janus,Lanata_FeSe_LDA+Gutz}, and it is the strongest when $M\pm 1$ electrons populate M orbitals (because this is the case with the strongest reduction of coherence among those in which the Mott insulator is disfavored), which is the case of stoichiometric Fe-based pnictides and chalcogenides, these materials bearing 6 electrons in 5 orbitals.

The complete phase diagram for a 5-orbital degenerate Hubbard model (with half-bandwidth D) as a function of filling and interaction strength
is reported in Fig. \ref{fig:5bands}.
This calculation is performed for a fixed ratio $J/U$ in the proper range for Fe-SC within this method\footnote{\label{foot:ratioJU}Indeed following the c-RPA estimates reported in Fig. \ref{fig:Miyake} one finds $J/U\simeq 0.12-0.16$. In order to properly implement this value in the semi-quantitative slave-spin mean-field approximation (SSMF), they have to be slightly enhanced. In practice in order to match the $U_c$'s for the Mott transitions for $J/U\simeq 0.15$ in DMFT with Kanamori interaction corresponds to $J/U\gtrsim 0.2$ in SSMF with density-density interaction (for $U_c$ in a 5 band model in DMFT see Ref. \cite{Lauchli_Werner_Krylov_5band}).}
and in the figure a square shows the realistic range for the Fe-SC, (filling from 5.5 to 6.2, and U from Fig. \ref{fig:Miyake} converted in units of the Hubbard model by matching the bare kinetic energies of the model and of the band-structures).

From this analysis it is clear that stoichiometric Iron superconductors are in some sense far from the Mott transition for 6 electrons in 5 orbitals, which happens at a much higher interaction strength (even if instead of $J/U$ one keeps $J$ fixed to the estimated value and raises only $U$). However these materials are nowhere near weakly correlated. Indeed even if far from the Mott transition at n=6, they are deep enough in the Janus phase, so that their quasiparticle weight is strongly reduced by Hund's coupling. The effect of pushing away the Mott phase makes that this is not a result of fine tuning, but is very solid to any variation of the physical parameters, be it intended both as an error bar on the theoretical estimates and as a variability among different materials. 

Indeed this simple model does not take into account the realistic band-structure of Fe-SC (like the crystal-field splitting of orbital levels, which partly contrast the effect of Hund's coupling) but as ab-initio calculations show it captures very well the physics, as far as the overall correlation strength is concerned.

%%%%%%%%%%%%%%%%%%%%%%%%%%%%%%%%%%%%%%%%%%
\begin{figure}[h]
\sidecaption
\includegraphics[width=6cm]{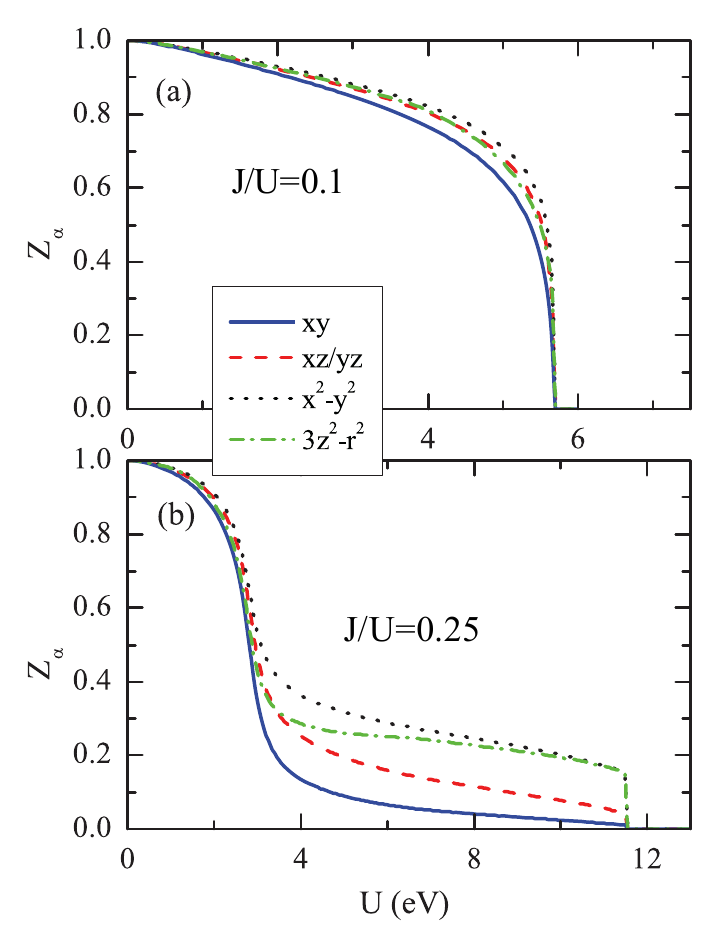}
\includegraphics[width=6cm]{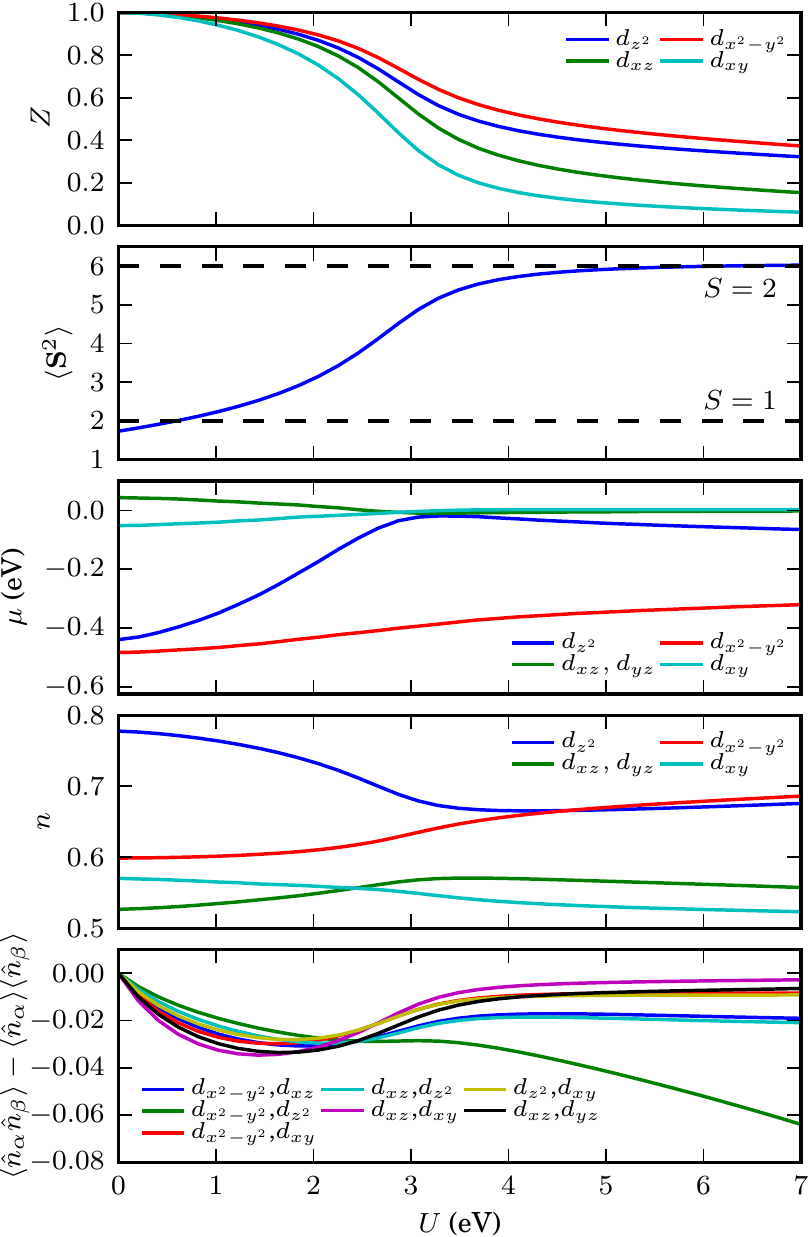}
\caption{Left: LDA+ Slave-spin calculation for LaFeAsO DFT band-structure for weak (upper panel) and strong (lower panel) Hund's coupling (adapted from Ref. \cite{YuSi_LDA-SlaveSpins_LaFeAsO}). The quasiparticle weight Z(U) (here resolved by orbital nature) assumes typical S-shaped curves of the Janus effect at strong Hund's coupling. Right: LDA+Gutzwiller analogous calculation for FeSe, from ref. \cite{Lanata_FeSe_LDA+Gutz}. It is to be noticed that the Janus phase is accompanied by the increase of inter-orbital spin correlations (i.e. a high-spin local moment is formed owing to the increase of J) and a suppression of the inter-orbital charge ones.}
\label{fig:Yu_Si_Lanata}   
\end{figure}
%%%%%%%%%%%%%%%%%%%%%%%%%%%%%%%%%%%%%%%%%

Realistic ab-initio calculations including the many-body correlations on top of DFT band-structures have indeed been performed from the outset (see the list of the different used methods in the introduction \cite{Yin_kinetic_frustration_allFeSC,Haule_pnictides_NJP,Shorikov_LaFeAsO_OSMT,Laad_SusceptibilityPnictides_OSM,Craco_FeSe,Aichhorn_FeSe,Ishida_Mott_d5_nFL_Fe-SC, Liebsch_FeSe_spinfreezing,Werner_122_dynU,Yin_PowerLaw, Misawa_d5-proximity_magnetic,YuSi_LDA-SlaveSpins_LaFeAsO,Yu_Si_KFeSe,Bascones_OSMT_Gap_halffilling,Ikeda_pnictides_FLEX,Lanata_FeSe_LDA+Gutz,Skornyakov_BaFe2As2_linearsusc}.) and fit very well into this global picture.

All of them regardless of the fine details point to a moderate to strong overall correlation strength. 
In Fig. \ref{fig:Yu_Si_Lanata} I report two of them, realized with methods numerically light enough to perform a full scan of the interaction strength (at fixed $J/U$, customarily).
These calculations realize, in specific cases, the general salient features of the Janus phase, and of Hund's metals:

\begin{itemize}
\item moderate to strong electronic correlations are realized (panel b on the left, top panel on the right), even far from the Mott insulator that is realized at very high interaction strength at this filling,
\item interorbital spin-spin correlations are strong, the high-spin state ($S=2$) is quickly predominant when inside the Janus phase (right, second panel from the top),
\item Hund's coupling favors the high-spin state by redistributing the orbital populations (right, second panel from below), by bringing the effective interacting orbital levels at roughly the same energy (right, third panel),
\item interorbital charge-charge correlations are suppressed (right, bottom panel).
\end{itemize}

A few further annotations are important at this point:
\begin{itemize}
\renewcommand\labelitemi{--}
\item Albeit at strong Hund's coupling the high-spin is realized, the crystal-field splitting of the orbital levels competes with it, favoring low spins. The low-to-high spin crossover happens at interaction strength quite near the actual realistic values for Fe-SC. This appears on the Z(U) curves as a very quick drop of coherence, which signals that the system enters the Janus/Hund's metal phase, in which the correlations are enhanced by the high-spin state. The position of this boundary depending on J, this explains the strong dependence on J of the coherence signaled in Ref. \cite{Haule_pnictides_NJP}.
\item Once the high spin state is realized one can consider the overall correlation strength described by the diagram in Fig. \ref{fig:5bands}. It is quite noticeable that the Mott transition at half-filling influences most part of the phase diagram. In particular, for an extended zone in the interaction range $U^{n=5}_c\lesssim U \ll U^{n=6}_c$ and for fillings around half, i.e. $4\lesssim n \lesssim 6$, the correlation strength is rather \emph{independent of both U and J}. The low coherence scale is rather set, for a quite extended range, by the doping from the Mott insulating state realized at half-filling, in a way that reminds the doped Mott insulator in the one-band case.
\item  The overall very reduced quasiparticle weight is reflected in a low coherence temperature of the metal. The phase just above such a coherence temperature was named in Ref. \cite{Werner_spinfreezing} "spin-freezing" phase, characterized by persistent in time spin-spin correlations and anomalous self-energies with low-energy power-law frequency dependence.
\item Finally one can notice that in the Hund's metal phase there is an increased tendency towards a differentiation of the correlation strength among the electrons in the various orbitals.  
\end{itemize}
In the following I will show that this differentiation of the correlation strength among orbitals is not the result of fine tuning, or of peculiarities of the electronic structure, but a solid character of the Hund-dominated metals, as first put forth in Refs. \cite{demedici_3bandOSMT, demedici_MottHund}. I will show that this differentiation is tied to the dominance of the half-filled Mott-insulator even on filling ranges  quite far from half and that this leads consistently to the coexistence of weakly and strongly correlated electrons in these materials, supported by a wide range of experiments.

\section{Orbital-selective Mott physics: experimental and \emph{ab initio} evidences}\label{sec:exp_abinitio_OSM}

Several experiments point in favor of a coexistence of multiple electronic components with different degrees of correlation/localization.

Multiple components in optical conductivity data in the Drude/MIR range\cite{Lucarelli_Optics_Co122, Wang_Optics_doped122}) as well as in magnetotransport\cite{Yuan_magnetoresistance_122_local-itinerant} reported for doped 122 pnictides were interpreted as a sign of more itinerant and more localized electrons coexisting, while ARPES studies\cite{Ding_Arpes_BaK, Yoshida_ARPES_KFe2As2} highlighted a strong Fermi-sheet dependence of the Fermi velocities.  
A strong orbital-dependence of the superconducting gap with hole-doping was pointed out in Ref. \cite{Malaeb_abrupt_gap_change_BaK122}, leading to a disappearance of only one gap in superconductive Ba${}_{0.4}$K${}_{0.6}$Fe${}_2$As${}_2$

In Iron chalcogenides NMR and EPR\cite{Arcon_NMR_EPR_FeSeTe_local-itinerant} and neutron scattering measurement\cite{Xu_neutrons_FeTeSe_local-itinerant} were intepreted as showing the presence of intrinsic local magnetic moments in the metallic non-superconducting phase, coexisting then with itinerant electrons.
ARPES\cite{Tamai_ARPES_FeSeTe_strongcorr} detected strong orbital differentiation and in particular stronger correlations for the xy orbital in  FeSe${}_{0.42}$Te${}_{0.58}$ and a study\cite{Yi_Shen_ARPES_OSMT_KFeSe} the intercalated chalcogenides A${}_x$Fe${}_{2-y}$Se${}_{2}$ (A=K, Rb) reported the disappearance of the band of xy-character when the temperature is increased above $\sim 150K$, thus signaling the orbital-selective Mott transition of the most correlated electrons. This was confirmed by THz spectroscopy\cite{Wang_Deisenhofer-OSMT_KFe2Se2}.

$K-\beta$ fluorescence XES measurements\cite{Gretarsson_LocalMoments} showed the presence of localized moments on the Fe 3d-shell due to the strong electronic correlations, thus illustrating the double nature of the electrons in the metallic phase. 

These scattered works within the contrasting evidences on the overall correlation strength (as briefly recalled in the introduction and summarized e.g. in Ref. \cite{Johnston_Review_FeSC}) could not  precisely clarify the role of electronic correlation and its selectivity, though.

More recently however, a main trend within the most studied family of doped BaFe$_2$As$_2$ was individuated in Ref. \cite{demedici_OSM_FeSC}, combining a theoretical study of the orbitally resolved correlation strength in these compounds, and a survey of the experimental mass enhancements as estimated by different techniques,  for the tetragonal paramagnetic metallic phase.

The compilation of such mass enhancements, from low-temperature specific heat, optical conductivity, ARPES and quantum oscillation measures, is reported in Fig. \ref{fig:Exp_OSM_FeSC} (technical details can be found in the supplementary online material of Ref. \cite{demedici_OSM_FeSC}).
What is found is a rapid increase with reducing filling of the mass enhancement estimated by the ratio of the measured to theoretical (as calculated from DFT)  Sommerfeld coefficient of the low-T linear contribution to the specific heat. Instead the reduction of the estimated Drude contribution within the optical conductivity spectrum does not quite follow, showing a more moderate mass enhancement rather saturating with hole doping.
%%%%%%%%%%%%%%%%%%%%%%%%%%%%%%%%%%%%%%%%%%
\begin{figure}[h]
\sidecaption
\includegraphics[width=7cm]{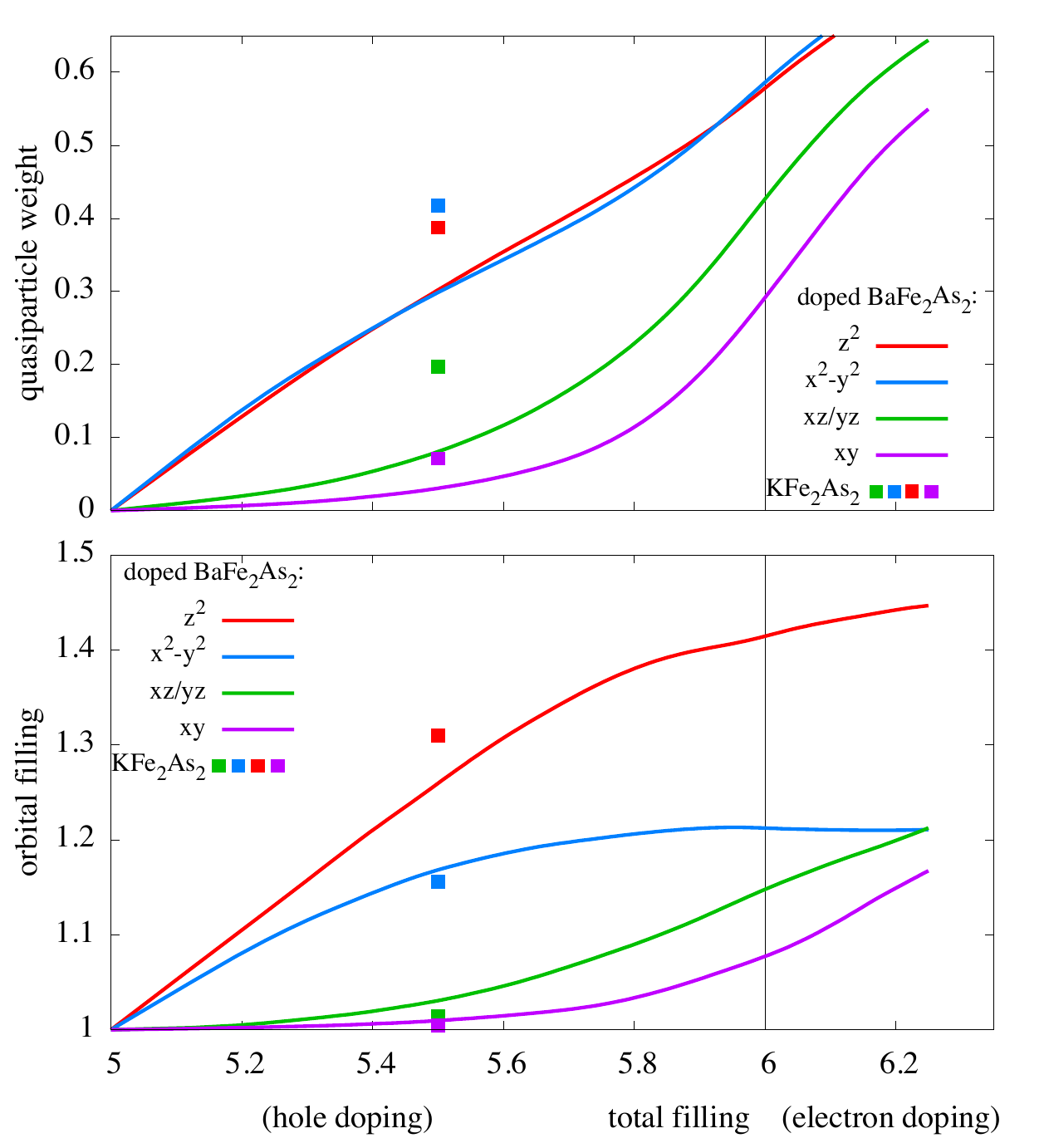}
\caption{Upper panel: orbitally-resolved quasiparticle weight as a function of the total conduction-band filling $n$ for doped BaFe$_2$As$_2$ (lines) and KFe$_2$As$_2$ (squares), as calculated with DFT+SSMF (in this mean-field context the quasiparticle weights are the inverse of the mass enhancements reported in Fig. \ref{fig:Theo_OSM_FeSC}). It is to be noticed that no sign of the commensuration is found at the stoichiometric filling $n=6$. A Mott insulator is realized instead for half-filled conduction bands ($n=5$). Lower panel: corresponding orbital populations. From  \cite{demedici_OSM_FeSC}.}
\label{fig:Z_norb}   
\end{figure}
%%%%%%%%%%%%%%%%%%%%%%%%%%%%%%%%%%%%%%%%%

This was interpreted in the light of theoretical calculations of orbitally-resolved correlation strength, performed within DFT+Slave-spin mean-field (Figs. \ref{fig:Theo_OSM_FeSC} and \ref{fig:Z_norb}), in the same work Ref. \cite{demedici_OSM_FeSC}. 
These show, in agreement with previous studies\cite{Ishida_Mott_d5_nFL_Fe-SC,Ikeda_pnictides_FLEX,Misawa_d5-proximity_magnetic} an asymmetry of the degree of correlation upon doping around the stoichiometric filling of 6 electrons/Fe. 
Indeed the theory indicates that correlations increase monotonically with reducing filling in the tetragonal paramagnetic metallic phase of the 122 materials, in a continuous trend that goes from the electron-doped to the hole-doped part of the phase diagram. 
Interestingly the correlation strength does not evolve identically for all electrons: those in the $t_{2g}$ orbitals (and in particular in the $xy$) are renormalized more strongly than those in the $e_g$'s. This differentiation of the mass enhancements grows with hole doping and culminates for the end member KFe$_2$As$_2$, for which they range from $\sim2.5$ (for the $e_g$'s) to $\sim14$ (for the $xy$).

This increasing spread among the electron effective masses is expected to result in a stronger renormalization of the Sommerfeld coefficient $\gamma/\gamma_{band}$ (where the $_{band}$ subscript refers to the unrenormalized value from band theory) compared to that for the Drude part of the optical conductivity $D_{band}/D$. Indeed the former is proportional to the total density of states at the Fermi level, which is a sum over the band (or orbital) index $\alpha$ of the bare contributions enhanced by factors $(m^*/m_{band})_\alpha$ (where $m_{band}$ is the bare band mass). The latter is instead a sum of contributions renormalized by the inverse factors $(m_{band}/m^*)_\alpha$. Thus $\gamma/\gamma_{band}=D_{band}/D$ only if the renormalization is the same for all orbitals (bands). In the opposite case of strong differentiation of the factors $(m^*/m_{band})_\alpha$, they will diverge from one another: as it happens for a series or a parallel of resistances, the Sommerfeld coefficient will be dominated by the heavier electrons, while the Drude part from the lighter ones. This is exactly what one observes in the experiments reported in Fig.\ref{fig:Exp_OSM_FeSC}.

This analysis is confirmed by ARPES and quantum oscillation measurements, which are band selective probes. Indeed the mass renormalizations of the different fermi surfaces estimated by these techniques are concentrated in the range $1.5\div3$ in the electron-doped compounds, while they spread more and more with hole doping, reaching extreme differentiations (in the range $\sim2\div 20$) in KFe$_2$As$_2$.

These consistent evidences strongly support the existence of weakly \emph{and} strongly correlated electrons in the whole phase diagram of the 122 family, the differentiation being tuned by the carrier doping and reaching extreme values at the end member with the lowest filling of 5.5 electrons/Fe.

%%%% MOTT
It is remarkable that the stoichiometric filling (6 electrons/Fe) does not represent a special point for correlations.
Half filling (5 electrons/Fe) is one, instead , because if the compound could be hole-doped to that point it would become Mott insulating\footnote{Here I refer to an ideal electrostatic doping, i.e. to a simple shift of the overall filling. Indeed the actual chemical doping is effectuated through atomic substitution, that modifies also the bare bandstructure. This is also true for KFe$_2$As$_2$, and indeed in Figs. \ref{fig:Theo_OSM_FeSC} and \ref{fig:Z_norb} the calculations for the actual DFT bandstructure of KFe$_2$As$_2$ are reported (squares) together with those for doped BaFe$_2$As$_2$ (i.e. in the so-called "virtual crystal approximation"). The two calculation differ mainly because KFe$_2$As$_2$ has a larger bare bandwidth. For both bandstructures however  a Mott insulator is realized at half-filling for the chosen values of $U=2.7eV$ and $J/U=0.25$ (cfr. Fig. \ref{fig:Miyake} and footnote \ref{foot:ratioJU}).}.
This is not actually possible, because the usual K substitution can only reach a filling of 5.5 electrons/Fe, for the end member KFe$_2$As$_2$ (however a very similar compound having a half-filled shell and similar electronic structure, BaMn$_2$As$_2$, is interestingly an insulator\cite{An_Sefat-BaMn2As2_ins}). Nevertheless, the calculations indicate that the influence of this Mott insulating state extends to the whole range of filling of interest for iron pnictides (i.e. up to and even beyond 6 electrons/Fe).

It is of great interest, in the view of the author, to show that the two previous aspects, the selectivity of correlations and the influence of the half-filled Mott insulator, are not specificities of the particular system under examination, but rather generic features of materials with conduction bands issued by nearly half-filled 3d shells and strong Hund's coupling, and can be understood in terms of the basic features of simpler models, which will be done in the next section.

\section{Orbital decoupling, the mechanism of selective Mottness}\label{sec:orb_dec}

We have shown in Section \ref{sec:overall_corr} that some of the prominent features found in the realistic calculations for Fe-SC are already present in a simple Hubbard model of 5 degenerate orbitals giving rise to 5 degenerate bands with featureless semi-circular densities of states. Indeed in this simple model (see Fig. \ref{fig:5bands}) a Mott insulator is realized at half-filling for interaction strengths in the realistic range for Fe-SC, and its influence extends over a large part of the phase diagram. 

We will analyze this model further here, and show that the origin of the extended (in doping) range of influence of the Mott insulator is the same than for the tendency to differentiated correlation strengths among the orbitals once their degeneracy has been removed. It is the emergent mechanism that we call "orbital decoupling".

\begin{figure}[h]
\sidecaption
\includegraphics[height=6cm]{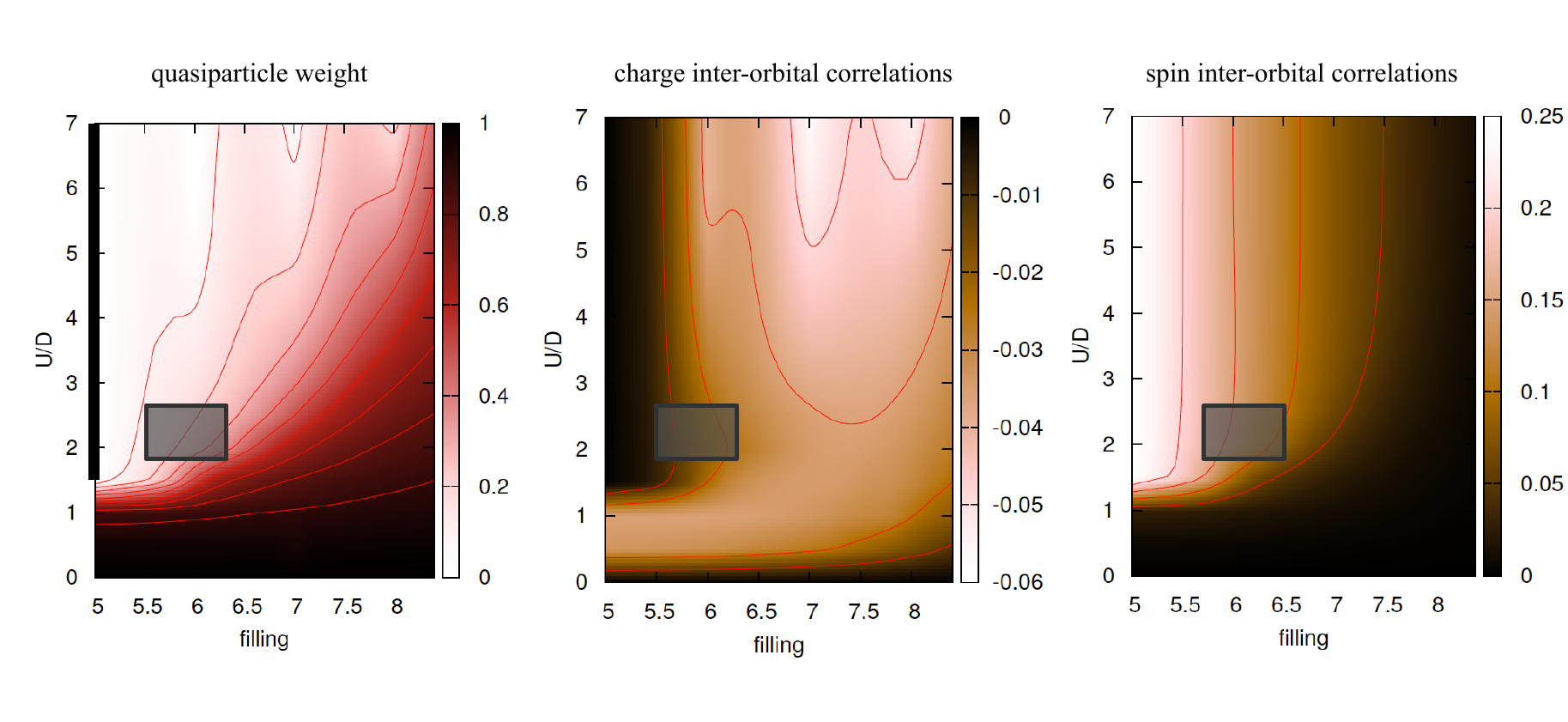}
\caption{Quasiparticle weight (left panel), inter-orbital charge charge (center) and ferromagnetic spin-spin (right) local correlations, for a 5-band degenerate Hubbard model at fixed $J/U=0.2$. The Mott insulator realized at half-filling is indicated by the black bar in the left panel. The grey square shows the realistic range of interaction and filling for Fe-SC (see also the caption in Fig.  \ref{fig:5bands}).}
\label{fig:Z_corr-half}   
\end{figure}

Indeed in Fig. \ref{fig:Z_corr-half} beside the quasiparticle weight (left panel, reporting the right half of Fig. \ref{fig:5bands}), the inter-orbital charge-charge (central panel) and ferromagnetic spin-spin (right panel) local correlations are reported, as a function of the interaction strength (at large $J/U$) and total filling for the fully degenerate 5-orbital Hubbard model. 
%%%%%%%%%%%%%%%%%%%%%%%%%%%%%%%%%%%%%%%%%%
%%%%%%%%%%%%%%%%%%%%%%%%%%%%%%%%%%%%%%%%%
What I want to single out here is the contrast between the monotonic behavior in doping of the spin-spin correlations, that show the progressive build-up of the local moment, in approaching the hig-spin ($S=5/2$) realized in the Mott insulator at half-filling, and the non monotonic behavior of charge-charge correlations. Indeed these show a large suppression in a wide area around the Mott insulating state, extending (in particular for the realistic interaction strength for FeSC indicated by the grey square) to and beyond the stoichiometric filling for FeSC of 6 electrons/Fe.

This emergent behavior is a sign of what has been called \emph{orbital-decoupling} in Ref. \cite{demedici_OSM_FeSC} (or band-decoupling in Ref. \cite{demedici_MottHund}), and is an implication of the high-spin Mott insulator realized at half-filling in presence of strong Hund's coupling. 
The suppression of the orbital susceptibility in the 2-orbital half-filled Mott insulator in presence of strong Hund's coupling had already been reported in Ref. \cite{Werner_Hund}. 
This behavior implies that the charge excitations in the different orbitals are largely independent, and remarkably this independence survives a considerable amount of doping.
Charge excitations are the basic object of Mott physics and tune the correlation strength in the proximity of a Mott insulator. Being independent in each orbital this gives way to independent tuning of the correlation strength in each orbital, once the perfect symmetry between the orbitals of the model is removed.

\begin{figure}[h]
\sidecaption
\includegraphics[width=7cm]{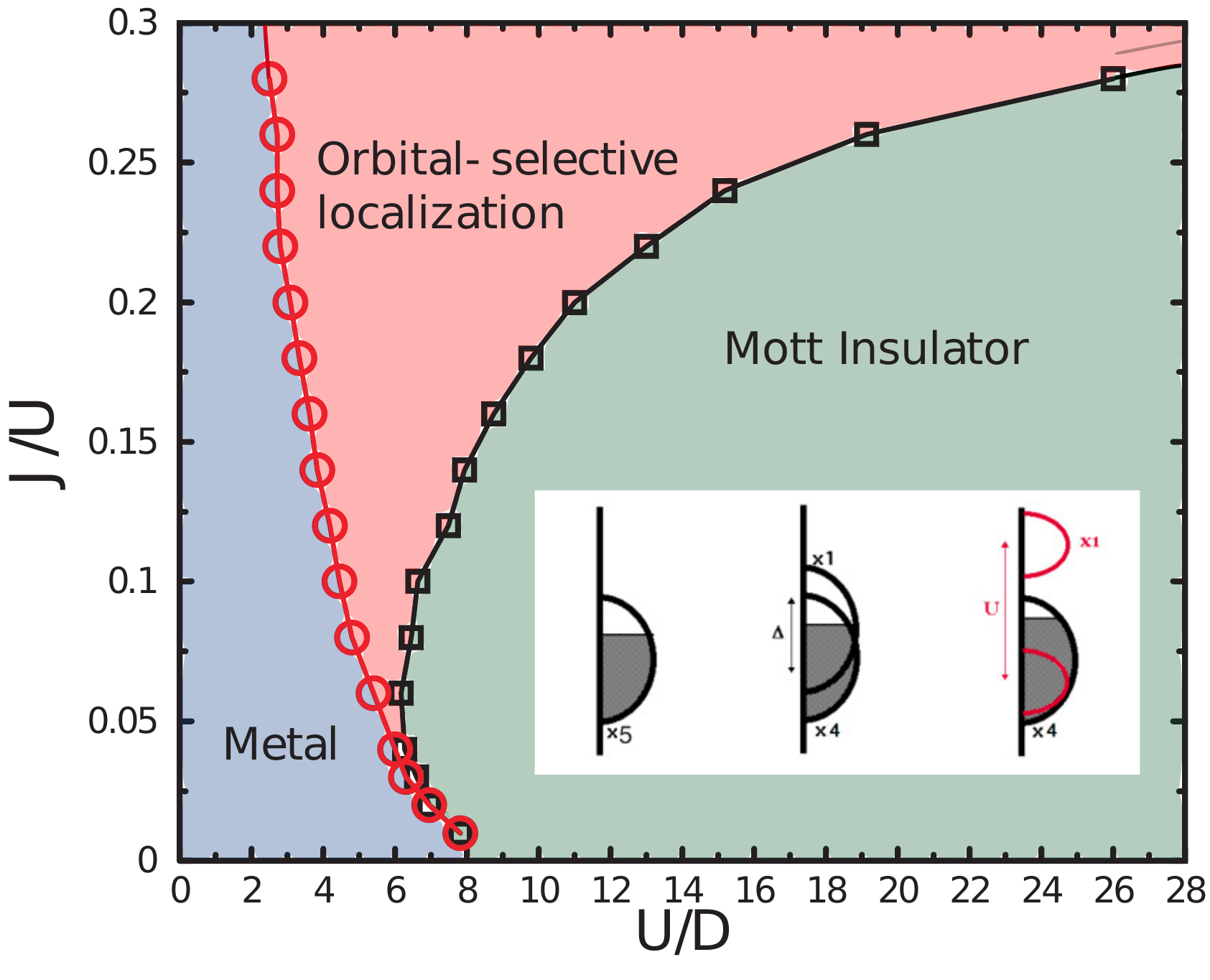}
\caption{Phase diagram of a 5-band Hubbard model (for $J/U=0.25$) in presence of a crystal-field (as sketched in the inset) lowering in energy 4 out of 5 orbitals. The total population is fixed to 6 electrons, while the crystal field is adjusted so that the upper orbital is kept half-filled. An orbital-selective Mott state is found, beyond a critical $J/U$, between a metallic phase at weak coupling and a Mott insulator at strong coupling. From Ref. \cite{demedici_Genesis}.}
\label{fig:5bands_OSMT}   
\end{figure}

Indeed if a crystal-field splitting is introduced in the previous model, where one orbital is kept half-filled, while the others are lowered in energy (remaining degenerate) and filled until reaching the overall stoichiometric population of 6 electrons in 5 orbitals, the phase diagram of the model becomes the one reported in Fig. \ref{fig:5bands_OSMT}: an orbital-selective Mott state is created, where the electrons in the half-filled orbital keep a gap to the charge excitations and form thus a Mott insulator, while electrons in the remaining orbitals delocalize and create a metallic state\cite{demedici_Genesis}.
%%%%%%%%%%%%%%%%%%%%%%%%%%%%%%%%%%%%%%%%%%
%%%%%%%%%%%%%%%%%%%%%%%%%%%%%%%%%%%%%%%%%

The one reported here is only one example of orbital-selective Mott transition (OSMT, the extreme case of selective differentiation in correlations, where electrons in subset of orbitals are localized while others remain itinerant) promoted by Hund's coupling found in many models in recent studies (see Section 6.6 of Ref \cite{Georges_Annrev} for a discussion and reference list).

\begin{figure}[h]
\sidecaption
\includegraphics[width=5cm]{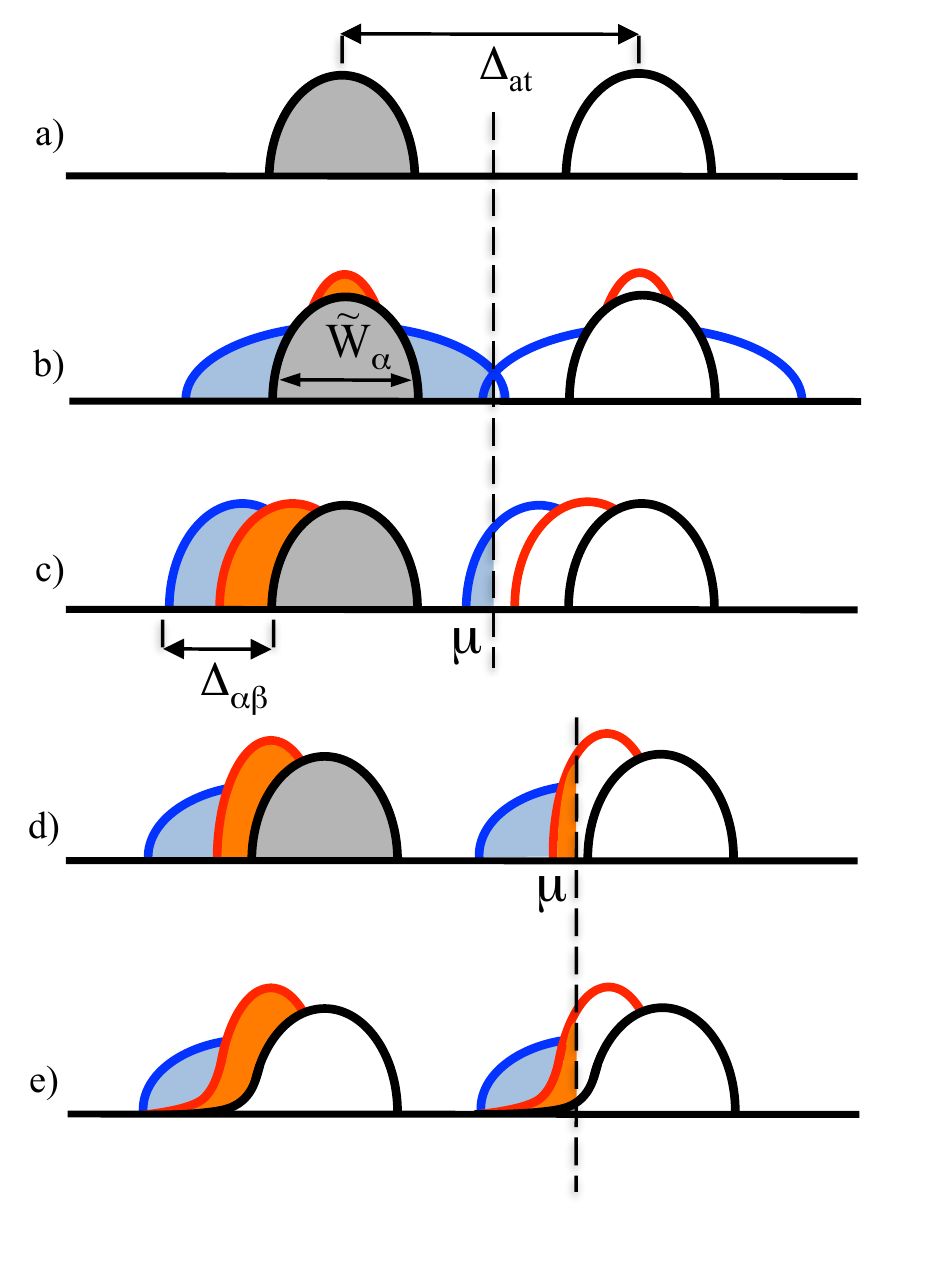}
\caption{Cartoon of a general mechanism for Hund's promoted orbital-selective Mott transitions and selective Mottness (see text). a) Spectrum of a degenerate half-filled Hubbard model; b) Orbital selective Mott phase triggered by a difference in bandwidths; c) Orbital selective Mott phase triggered by a lifting of degeneracy of the orbital energies due to crystal-field splittings; d) General situation with both the previous effects combined, OSM phase due to doping; e) Onset of hybridization compared to d), transforming an orbital-selective Mott phase in a metallic phase with orbital-selective Mottness. }
\label{fig:OrbDec_cartoon}   
\end{figure}

The majority of these can be understood using a simple cartoon construction of the orbitally-resolved spectral functions, as shown in Fig. \ref{fig:OrbDec_cartoon}.
From top to bottom the spectral functions are schematically constructed following a simple procedure:
\begin{description}

\item [a)] One starts from a caricature of a half-filled multi orbital Mott insulator. Following the atomic spectrum two "Hubbard bands" exist at a distance set by the atomic Mott gap $\Delta_{at}$ ($=U+(M-1)J$, for an M-orbital Hubbard model at half-filling). The strong Hunds coupling ensures that the orbitals are individually half-filled in the Mott insulating state.  

\item [b)] these Hubbard bands disperse independently for  each orbital $\alpha$, on a range set by the bare bandwidth $\tilde{W}_\alpha$ associated to each orbital (we suppose that the hopping integrals are diagonal in the orbital index for the moment). This independent dispersion happens thanks to the decorrelation of the charge excitations due to the orbital decoupling mechanism. An early strong-coupling argument explaining the independence of Hubbard bands was given in Ref. \cite{Koga_OSMT}. 

Applying an orbital-selective version of the Hubbard criterion, some orbital can have overlapping Hubbard bands, while others can have an open gap, independently. Depending on the different combinations of open or closed gap dues to the bandwitdths associated to the different orbitals, the system can be in the metallic, orbitally-selective Mott or Mott insulating state, at half-filling. This is the case of the "standard model" for OSMT studied in many works (a two-orbital Hubbard model with different bandwidth - see Ref. \cite{Inaba_dopedOSMT} for an extensive list of references).

\item [c)] The vanishing orbital susceptibility implies a rigidity of the Hubbard bands upon orbital energy shifts. Thus the bands are shifted relatively to one another by the crystal-field splittings $\Delta_{\alpha\beta}$ if present. This alone, is another way to realize an OSMT\cite{demedici_3bandOSMT,Lauchli_Werner_Krylov_5band,demedici_MottHund}, if an orbital is shifted enough in energy in order for one of his Hubbard bands to reach zero energy (while keeping an open gap away from it), while the rest of the system has a still an open gap at the chemical potential. Among the metallic bands it is found that the correlation strength is set by the proximity of each orbital to individual half-filling\cite{demedici_MottHund}. The behavior of each orbital reminds that of a single-band doped Mott insulator in which the quasiparticle weight grows linearly with doping (see section \ref{sec:overall_corr}).

\item [d)] Panel d) in the Fig.  \ref{fig:OrbDec_cartoon} depicts the general case, where both bandwith differences and crystal field splittings contribute to the generation of displaced independent gaps for the different orbitals. Beside that mentioned in c) a general rigidity applies to the whole spectrum due to the vanishing compressibility of the Mott insulator. Thus the chemical potential $\mu$ can be moved and the whole spectrum shifts rigidly. Hence $\mu$ can lie within the Hubbard bands of some orbitals, while still being in the gap for the remaining orbitals. In this situation the system is necessarily doped away from half-filling and in an orbitally-selective Mott state\cite{Koga_OSMT,Werner_Hund,Jakobi_dopedOSMT, Jakobi_OSMT_cfs}. After a certain critical doping, when the chemical potential exits the last open gap, the system recovers a normal metallic phase\cite{demedici_3bandOSMT}.

\item [e)] Finally upon onset of hybridization (local and/or non-local hoppings) between the different orbitals it has been shown that the present cartoon is slightly smoothed (the gaps can be rounded in pseudogaps, or replaced by a heavy-fermionic metallic phase with very low compressibility replacing the incompressible plateau in the orbital resolved spectral functions\cite{Koga_OSMT_2005,demedici_Slave-spins,Winograd_pseudogap}), but the general structure of this cartoon is preserved. 

\end{description}
%%%%%%%%%%%%%%%%%%%%%%%%%%%%%%%%%%%%%%%%%%
%%%%%%%%%%%%%%%%%%%%%%%%%%%%%%%%%%%%%%%%%
A word of caution concerning this stylized general cartoon that I have given of Hund's dominated doped Mott insulators is in order. It is intended to describe the gross features of the orbital-resolved spectra (mainly the location and width of the gaps and the structure of the spectrum close to the gaps, not the spectrum at high energies) and the main changes of the local physics under tuning of the most important knobs in the system: chemical potential, crystal-field splitting, bandwidths associated to the different orbitals (the last is intended as the effective quantity $\tilde{W_\alpha}$ defined in Section \ref{sec:overall_corr}, which, if not coincident with the bare $W_\alpha$, should at least scale with it). It neglects all specificities coming from the k-space structure of the bands (nesting, Van-Hove singularities, direct or indirect gaps in the bandstructure, etc.).
Also it is not expected to be valid too far from the half-filled Mott insulator, in the sense intuitively defined by Fig. \ref{fig:5bands} and discussed at the end of Section \ref{sec:overall_corr}.
Finally, as discussed for the crystal-field in section \ref{sec:overall_corr}, also hybridization, if large enough, can reduce the effect of Hund's coupling, and even bring the system to a low-spin state. The cartoon given above, and more generally the whole analysis given in this chapter, are founded on the assumption that Hund's coupling dominates over both crystal-field splittings and hybridization. These are then treated as perturbations of the picture given for degenerate and non-hybridized orbitals, and when this hierarchy is inverted the present picture does not apply.

It is easily checked that most of the OSMT studies can be interpreted in the light of the cartoon given above. Among the many aspects that one can highlight, it is quite revealing for instance that the orbital-resolved spectra in models with featureless bands are quite similar among them, and simply shifted from one another in presence of crystal field splitting\cite{demedici_3bandOSMT, demedici_MottHund}) or different only in the width of the Hubbard bands when the gaps are opened in models with no crystal-field splitting but different bandwidths\cite{Koga_OSMT,Koga_OSMT_2005}.

\section{Back to realism: FeSC and two 'wrong' (yet instructive) calculations}\label{sec:realistic_Orb_dec}

The last snapshot of Fig. \ref{fig:OrbDec_cartoon} is the identikit of the situation found in FeSC in Ref.\cite{demedici_OSM_FeSC}.
Indeed this can be seen quite explicitly\footnote{The best discussion of the relevance of the proposed cartoon for the physics of FeSC would be to calculate explicitly the spectra which are meant to be stylized by the cartoon. However, detailed, reliable, low-temperature orbitally-resolved real-axis spectra are not yet easily obtained by state-of-the art DFT+DMFT techniques. We thus rely to the integrated spectral weight giving the orbital populations, much more easily and reliably available in SSMF (as well as in DMFT).} 
in the lower panel of Fig. \ref{fig:Z_norb}, where the orbital populations $n_\alpha$ as a function of the total filling are shown. These are quite obviously the outcome of a spectrum like the one sketched in the lowest panel of Fig. \ref{fig:OrbDec_cartoon} (the populations being obtained by integration of the orbitally-resolved spectral densities up to the chemical potential).

\begin{figure}[h]
\sidecaption
\includegraphics[width=9cm]{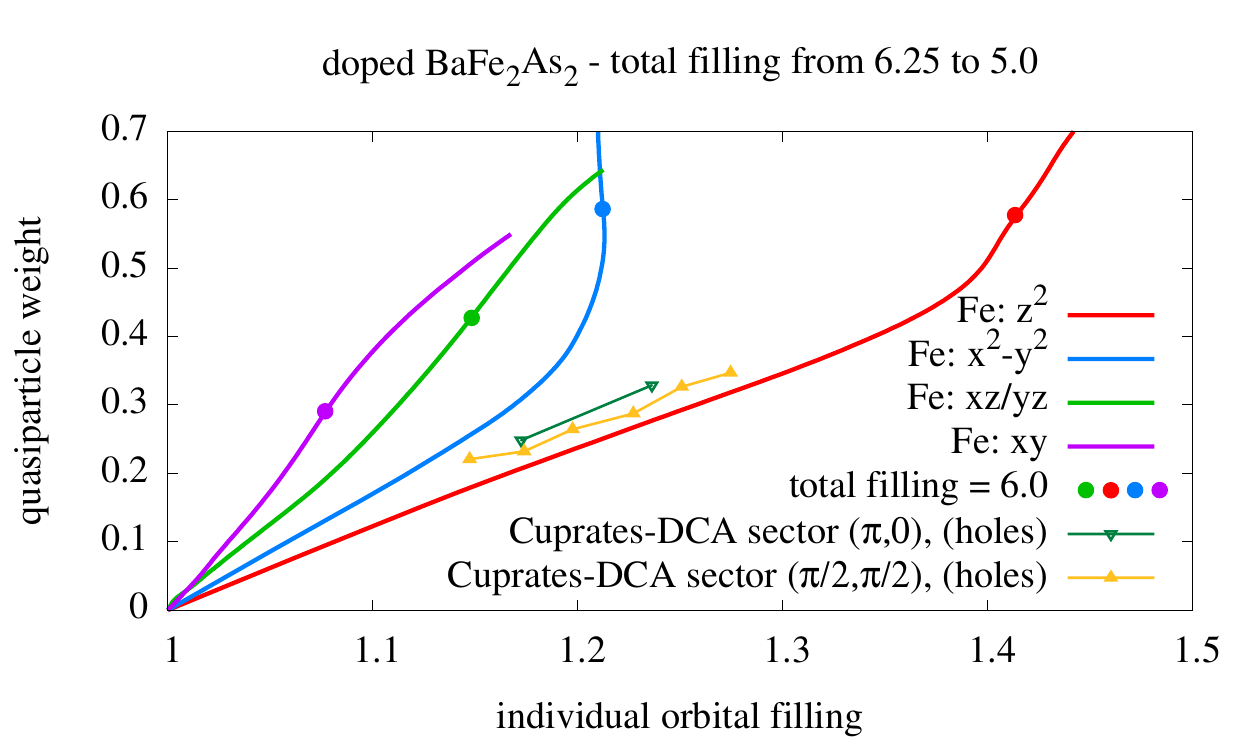}
\caption{Quasiparticle weight for electrons in each orbital as a function of the respective orbital population, as calculated within DFT+SSMF for  BaFe$_2$As$_2$. The striking linear behaviour shows that each orbital behaves as an individual doped Mott insulator. Comparison with an analogous decoupling found in the DCA calculations for the bidimensional Hubbard model from Gull et al. Ref. \cite{Gull_DCA_k-space_selec} is reported (see text). From Ref. \cite{demedici_OSM_FeSC}.}
\label{fig:OrbDec_BaFe2As2}   
\end{figure}

Moreover a very revealing feature is the fact that the orbitally resolved quasiparticle weights ($Z_\alpha$, upper panel of Fig. \ref{fig:Z_norb}), mirror this very trend. Indeed
when plotting, for each orbital, the quasiparticle weight as a function of the population of that orbital $Z_\alpha(n_\alpha)$, one finds a remarkably linear behavior, as shown in Fig. \ref{fig:OrbDec_BaFe2As2}. 
%%%%%%%%%%%%%%%%%%%%%%%%%%%%%%%%%%%%%%%%%%
%%%%%%%%%%%%%%%%%%%%%%%%%%%%%%%%%%%%%%%%%
This shows clearly that the correlation strength for the electrons in each of the orbitals is almost solely set by the proximity of that orbital to half-filling, and thus that each orbital behaves quite independently as an \emph{individual doped Mott insulator} in which, as reminded earlier on, the mean field Z is proportional to the doping from half-filling.
In particular it is found that for stoichiometric BaFe$_2$As$_2$, the $t_{2g}$ orbitals are already deeply in this Z-linear regime of orbital-selective Mottness (while the $e_g$ orbitals will enter it eventually, upon hole-doping). 

This selective Mottness can be regarded as an emergent phenomenon in that  the direct effect of the specific microscopic features of the model - such as the starting orbital populations, or crystal-fields, or the size of specific hopping integrals, as determined by the geometry of the lattice and symmetry of the orbitals - on the correlation strength found in each orbital is quite hard to disentangle.
However the relationship with the "intermediate" quantity (meaning that it cannot be independently tuned, but it is rather determined by all the previous factors combined in a non-trivial way) that is the final orbital population (i.e. as found in the fully interacting system) is instead direct, and easily shown to be the most relevant one. 
Moreover this happens upon entering the zone of reduced inter-orbital charge correlation near the Mott insulating state. These correlations are a non monotonous function of U and J (at fixed $U/J$), as shown already in the degenerate model (Fig. \ref{fig:Z_corr-half}) and confirmed explicitly in the realistic calculations (see supplementary material of Ref. \cite{demedici_OSM_FeSC}). This surprising reduction of these charge correlations upon increase of the interaction strength is a direct signal of the complexity involved in the relationship between orbital decoupling and microscopic couplings. 

Within this reverse-engineered view of the electronic correlations of FeSC many features found  in ab initio calculations (and confirmed by the analysis of experiments of ref. \cite{demedici_OSM_FeSC} and reported in this chapter) are quite naturally understood, such as:
\begin{itemize}
\renewcommand\labelitemi{--}
\item the asymmetry of the phase diagram around the total filling $n=6$, which clearly in the presently exposed rationale does not represent a special point, all physical quantities evolving monotonously through it, in the paramagnetic phase;
\item the increase of  correlations with hole-doping, since all orbital fillings (and thus the total density) are moving towards half-filling, where a Mott insulator is realized;
\item the extended influence of the half-filled Mott insulator, even up to the stoichiometric filling $n=6$ (and beyond). Even if globally doped one electron away from half-filling (that usually is considered "far"), the five orbitals are decoupled and then influenced by their individual populations. These are on the average only 20\% away from half-filling \cite{Ishida_Mott_d5_nFL_Fe-SC,demedici_OSM_FeSC} (and some actually just a few percents away from it). In this sense these materials can rather be considered "near" the n=5 Mott insulator;
\item the stronger correlations of electrons with character mainly of the $t_{2g}$ orbitals, and in particular the $xy$ ,which is the one systematically found the closest to half-filling in calculations. The $e_g$ orbitals remain in all calculations much more filled and are found much less correlated.
\end{itemize}

This last point calls for further analysis however. 

Naturally one would like to relate the stronger correlations found in the $xy$ orbital to the bare parameters of the bandstructure, and also identify the reason for their ubiquity. 

In ref. \cite{Yin_kinetic_frustration_allFeSC} the attention has been drawn to the smallness of the diagonal hopping integral for the xy orbitals (see table \ref{fig:main_hops}), due to the reciprocal cancellation of direct and ligand-mediated hopping amplitudes between these orbitals.
However it can be shown that this is not enough to determine the stronger correlations, by performing a calculation for a "wrong" bandstructure for BaFe$_2$As$_2$ in which all the hoppings are maintained unaltered, but the bare energy of the $xy$ orbitals is artificially lowered (keeping the total population fixed to 6 electrons/Fe). Predictably the population of the $xy$ orbital is found to increase, while the one of the other orbitals moderately decreases. In the Hund's metal regime (i.e. for $U~2.7eV$ and larger, for this bandstructure) the quasiparticle weights are found to follow the orbital populations, and thus they increase for the $xy$ orbitals and decrease for the other orbitals. This is the essence of the orbital-decoupling mechanism because, as illustrated by the cartoon given in this section, the orbitally-resolved spectra shift pretty rigidly and independently from one another. Moreover the orbitally-resolved quasiparticle weight are linear functions of the respective orbital populations, and are modified accordingly.
In this example when the bare $xy$ energy is moved from $\sim 0.1eV$ above (as in the realistic case, see table Fig. \ref{fig:Crystal_fields}) to $\sim0.1eV$ below the one for the $xz/yz$, the two orbital populations cross and so does the degree of correlation. For even lower energies the $xy$ have weaker correlations strength than the $xz/yz$ despite all the hoppings and the rest of the energies being unaltered.
%%%%%%%%%%%%%%%%%%%%%%%%%%%%%%%%%%%%%%%%%%
\begin{figure}[h]
\sidecaption
\includegraphics[width=6cm]{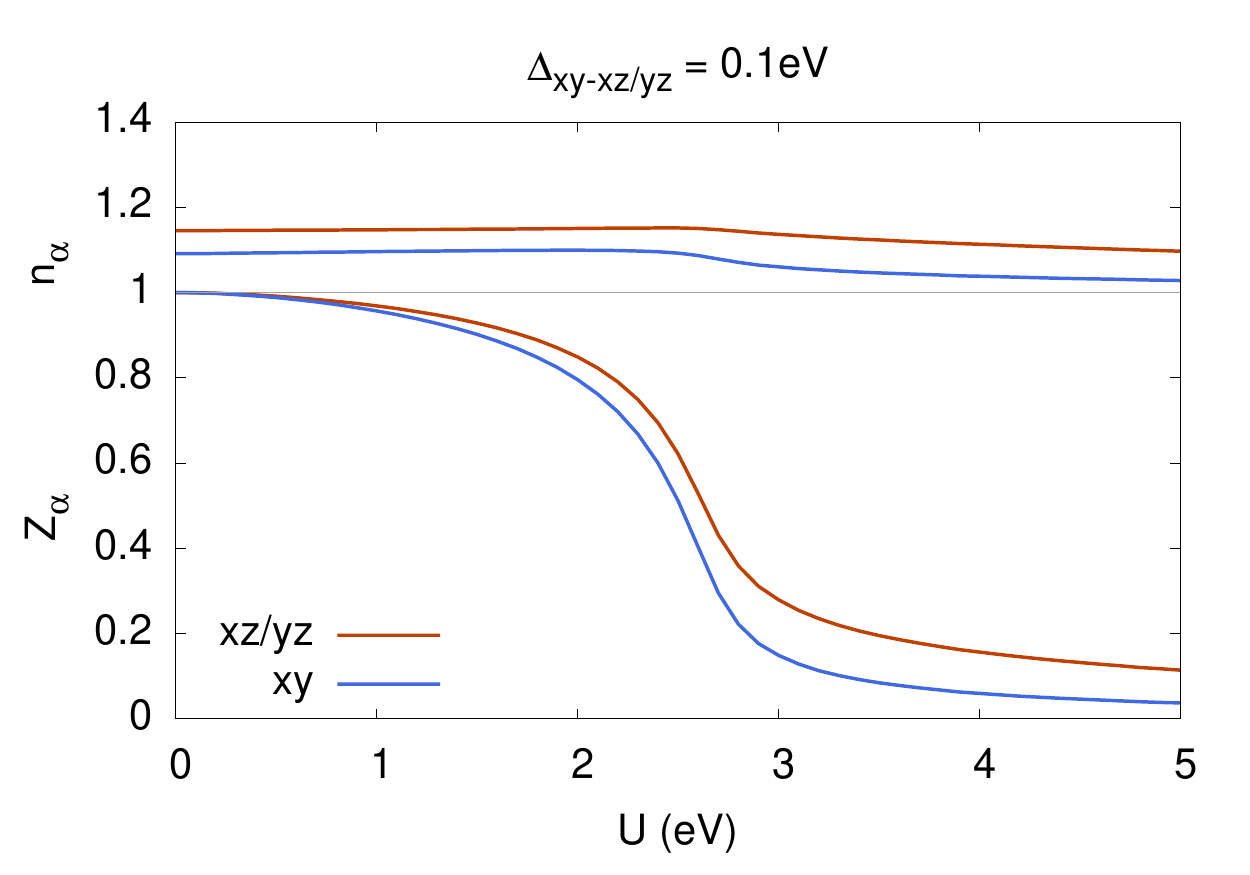}
\includegraphics[width=6cm]{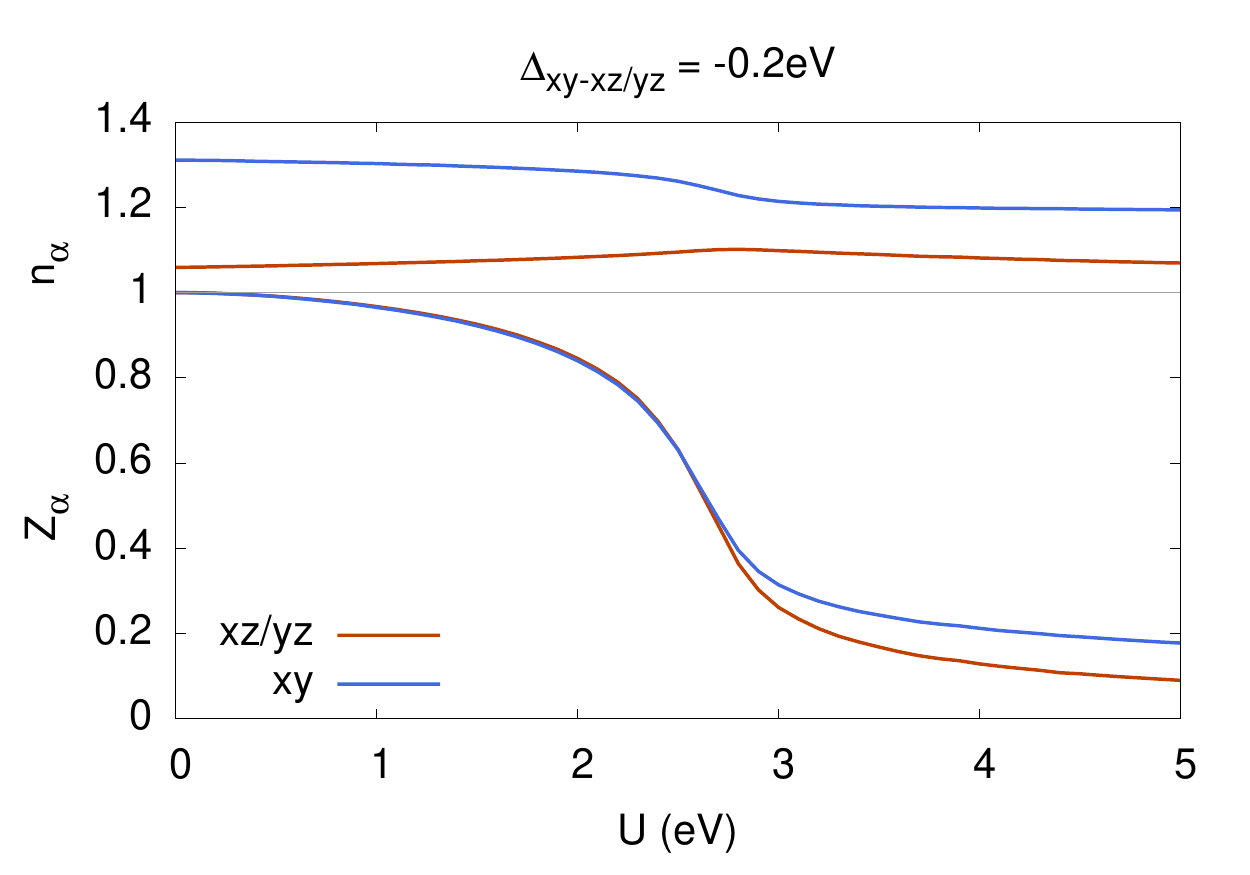}
\caption{Quasiparticle weight (lower curves) and populations (upper curves) for the $t_{2g}$ orbitals for the BaFe$_2$As$_2$ bandstructure (left, where the crystal-field splitting between $xy$ and $xz/yz$, $\Delta_{xy-xz/yz}\sim 0.1eV$) and for an altered bandstructure where the $xy$ energy has been artificially lowered so that  $\Delta_{xy-xz/yz}\sim - 0.2eV$ and all the other hoppings and energies has been conserved (J/U=0.25). This clearly shows that in the Hund's metal regime (here for $U\gtrsim 2.7$ eV) the correlation strength is mainly set by the distance of each orbital from half-filing, despite the fact that in both cases the nearest-neighbor hopping $t_{xy-xy}$  is much smaller than $t_{xz-xz}$ (see table \ref{fig:main_hops}).}
\label{fig:BaFe2As2_wrong_calc}   
\end{figure}
%%%%%%%%%%%%%%%%%%%%%%%%%%%%%%%%%%%%%%%%%

Another instructive "wrong" calculation can be done by modifying, in the original BaFe$_2$As$_2$ bandstructure only the diagonal $xy-xy$ hopping, by putting it equal to the $xz/yz-xz/yz$ one. Again in the Hund's metal phase the $xy$ orbitals end up being slightly closer to half-filling than the $xz/yz$, and consistently slightly more correlated. 
This second calculation is certainly less clear-cut than the previous one, since there are many other hoppings that can differentiate the kinetic-energy content of the different $t_{2g}$ orbitals, but is at least consistent with the general point that we are making that it is not the hopping structure alone that determines the correlation differentiation in a Hund's metal.

Nevertheless the smaller $xy$ hopping does play a role here. 
Indeed based on what I have just exposed one may think that the correlation strengths are simply tuned by the bare crystal-field splittings: indeed the result for BaFe$_2$As$_2$ reported in the lower panel of Fig. \ref{fig:Z_norb} follows the bare orbital energies reported in table in Fig. \ref{fig:Crystal_fields}. 
However this is not true for other FeSC, like KFe$_2$As$_2$ or the chalcogenides, where the non-interacting orbital energies are inverted, as visible in the same table. Indeed for all these materials the correlation end up being stronger in the $xy$ orbital, irrespectively of its energy being higher or lower than the $xz/yz$. This can even lead the $xz/yz$ being closer to half-filling than the $xy$ in the non-interacting system. However as shown in the second lowermost left panel in Fig. \ref{fig:Yu_Si_Lanata}, the interactions lead to a population crossing and within the Hund's metal the $xy$ indeed lies closer to half-filling, and the correlation strength is set accordingly. This is consistent with the $xy$ orbital having a smaller Hubbard band (due to the smaller bare hopping), in the view outlined by the orbital-decoupling cartoon of Fig. \ref{fig:OrbDec_cartoon}.

Thus neither the order in the bare orbital energies nor a hierarchy in the hopping amplitudes are sufficient to determine the correlation strengths in the various orbitals. 
What we highlight as the most solid trend is, once again, the final (i.e. in the fully interacting system) orbital population. Within the Hund's metal regime, thanks to the orbital-decoupling mechanism, the correlation strength scales linearly with it, and one can expect that the hierarchy of correlation strengths follows the one in the population, as we have found in all the performed calculations thus far (see also the supplementary material of Ref. \cite{demedici_OSM_FeSC}).

 Even this is not a rigorous result however. Indeed even if within the orbitally decoupled regime the  $Z_\alpha(n_\alpha)$ are linear, the slope is not universal and has some variation among the orbitals, as clearly shown in Fig. \ref{fig:OrbDec_BaFe2As2}. An exact understanding of the origin of the different slopes has not been reached as of today. There are indications that they are sensitive to the bare crystal-field splittings. This can be rigorously shown in a two band toy model (see Appendix). This is also consistent with the realistic cases of FeSC that we studied up to now (as reported in Ref. \cite{demedici_OSM_FeSC} and its supplementary material): the slopes go from the steepest for the orbitals with higher orbital energies to the least steep for the low-lying ones. This implies for example that the $Z_{xy}(n_{xy})$ has the steepest slope in LaFeAsO and BaFe$_2$As$_2$, while   $Z_{xz/yz}(n_{xz/yz})$ is steepest in the chalcogenides. 
 
If a proper rescaling of the slopes with the bare orbital energies could be worked out, the dependence of the correlation strength on the individual orbital filling would be universal and a rigorous result in the Hund's metallic phase. 

The take home message, nevertheless is that the most relevant quantity that sets the individual correlation strength in each orbital in the Hund's metal phase is the interacting distance in population from half-filling of that orbital. This is an emergent behaviour, as said in the introduction, because the interacting orbital populations are an "intermediate" quantity, set in a non-trivial way by the microscopic features of the system.
 
\section{Conclusions}\label{sec:conclusions}

In this chapter I hope to have given enough evidence pointing at the coexistence of strongly and weakly correlated electrons in the conduction bands of Iron-based superconductors.
This evidence is both experimental (from a lot of data available in the literature), and theoretical (from  calculations available in the literature, from previous work of the author and collaborators, and from some new material provided in this chapter).
Moreover it is well grounded in the fundamental emergent behaviour found in a Hund's metal in proximity of the half-filled Mott insulator arising for realistic values of the interaction strength that is \emph{orbital-decoupling}. The understanding of this basic mechanism helps understanding the reason for the observed correlation strength and also the action of the main physical knobs like hoppings and crystal-field splitting, even if it is found that the fundamental variables are the orbital populations of the interacting system.

A couple of last  remarks are in order.

Some aspects of orbital decoupling are still not understood in detail. Besides the aforementioned reason for the different slopes of the curves $Z_\alpha(n_\alpha)$, I would also highlight the role of hybridization, which is certainly depressed by Hund's coupling but essential, for instance, in inducing the heavy-fermionic behavior found at strongly hole-doped 122 FeSC. In chalcogenides, and in particular FeTe, hybridization seems instead incapable to prevent an actual orbital-selective Mott state, that indeed seems to be found experimentally\cite{Fobes_Zaliznyak-FeTe_OSMT}.

Also, we have well characterized and reverse-engineered the behaviour of FeSC, and in general of Hund's metals, once within the Hund's metal phase.
However the 1111 and 122 FeSC (those with the highest superconducting critical temperatures) lie most probably \emph{at the border} between the normal and Hund's metal phase. 
There may be specificities to be attributed to this particular position in the phase diagram, on the verge of the Hund's metal phase, which are still to be clarified.

Finally some very important issues have been left out of the scope of this chapter. It is worth citing that we have focused on the properties of the paramagnetic normal tetragonal phase, so that no discussion specific to the magnetic transition and to the structural one, as well as those pertaining to the superconducting phase transition and pairing mechanism has been tackled here.

Lastly I want to mention that I have not entered in the possible parallels that can be traced between the physics of FeSC and that of Cuprates. I want to mention nevertheless that based on the emergent physics of orbital decoupling a common phase diagram revolving around the half-filled Mott insulator and the ensuing (selective) Mottness has been put forth in our work Ref. \cite{demedici_OSM_FeSC}, where we have joined the present discussion of orbital decoupling in FeSC with a similar analysis of the data by Gull et al.\cite{Gull_DCA_k-space_selec}. 
We have highlighted that the selective Mottness found by Gull et al. in the bi-dimensional Hubbard model used for Cuprates and studied with the Dynamical Cluester Approximation, seems to stem from an orbital-decoupling mechanism very similar to that discussed here, only applying to the different areas (nodal, antinodal) of the Brillouin zone. Indeed when plotting the nodal/antinodal quasiparticle weight as a function of the fraction of doping that can be attributed to the nodal/antinodal area (see supplemental material of Ref. \cite{demedici_OSM_FeSC}) a slope very similar to the ones found for FeSC is obtained (see Fig. \ref{fig:OrbDec_BaFe2As2}), pointing to a very similare orbital-deoupling mechnism between FeSC and Cuprates.

\begin{acknowledgement} 
This chapter is heavily founded on Ref. \cite{demedici_OSM_FeSC}, a work that was performed with Gianluca Giovannetti and Massimo Capone, to whom I am indebted.
The realistic bandstructures discussed in this chapter are those used in Ref. \cite{demedici_OSM_FeSC}, and have been calculated by G. Giovannetti.
\end{acknowledgement}

\section*{Appendix: the slope of the linear $Z_\alpha(n_\alpha)$ in the orbital decoupling regime}
\addcontentsline{toc}{section}{Appendix: the slope of the linear $Z_\alpha(n_\alpha)$ in the orbital decoupling regime}

The direct proportionality of the quasiparticle weight to the individual orbital population when approaching the Mott insulator is the main evidence of the orbital decoupling mechanism induced by Hund's coupling in the models for FeSC. However it can be noticed from Fig. \ref{fig:OrbDec_BaFe2As2} that the the slope of the linear behaviour is not universal.

In order to have an indication on how the bandstructure determines the slopes of the linear $Z(n)$ for each orbital, I have performed a simplified analysis on a Hubbard model with two bands of equal half-bandwidth D, (with hopping integrals as appropriate for a doublet in a cubic environment, see e.g. Ref. \cite{Poteryaev_Cfs_Hyb_Mott}), slightly hybridized (by an interorbital hopping V=0.05D) and split by a crystal field $\Delta$. The Kanamori density-density for of the interaction is used, with U=3D and J=U/4 and the model is solved within slave-spin mean-field. 

A Mott insulator is found at half-filling and strong orbital differentiation in the mass enhancements for a large region of doping around it: as expected, for electron-doping (hole-doping being the same, for particle-hole symmetry), for a small crystal-field splitting the upper orbital is closer to individual half-filling and more correlated than the lower one. This situation becomes extreme very close to half-filling and an orbital-selective Mott transition takes place eventually (albeit retarded by the hybridization - unlike the 5-orbital case in the regime relevant for FeSC, where the hybridization actually prevents the OSMT from happening). This happens when the upper orbital reaches individual half-filling and has a Mott gap at the chemical potential while the lower band remains metallic until global half-filling, where it becomes Mott insulating too. 

%%%%%%%%%%%%%%%%%%%%%%%%%%%%%%%%%%%%%%%%%%
\begin{figure}[h]
\sidecaption
\includegraphics[width=9cm]{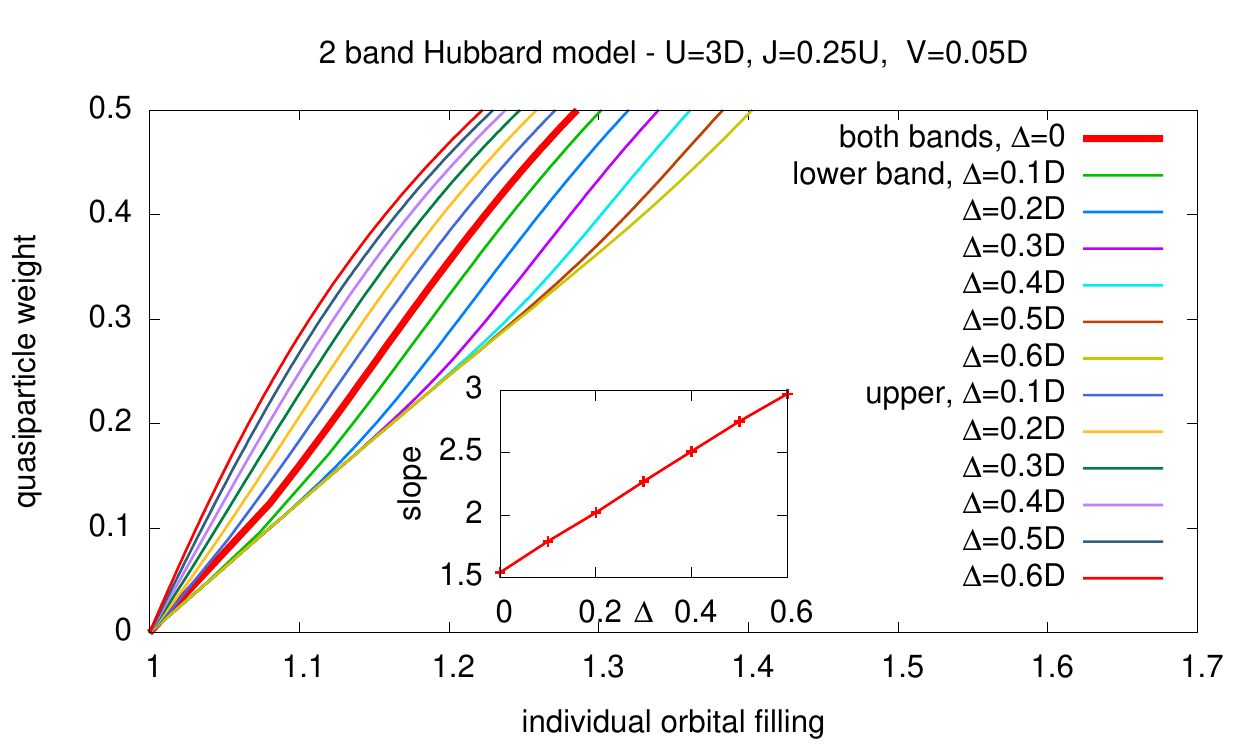}
\caption{Orbitally-resolved quasiparticle weight as a function of the respective orbital population in a two band Hubbard model with $U=3D$ and $J/U=0.25$ and a light hybridization $V=0.05D$, for several values of the crystal-field splitting $\Delta$ between the two orbital energies. The linear behavior typical of the orbital-decoupling near the Mott insulating state is indeed found in both orbitals. Inset: the slope of the steeper curve (corresponding to the orbital higher in energy) in the electron-doped case scales with the \emph{bare} crystal-field splitting.}
\label{fig:OrbDec_2band}   
\end{figure}
%%%%%%%%%%%%%%%%%%%%%%%%%%%%%%%%%%%%%%%%%

The curves $Z_\alpha(n_\alpha)$ are indeed linear (see Fig.\ref{fig:OrbDec_2band}) and the slope is steeper for the upper band. Thus at a given orbital population, the mass enhancement is heightened by the presence of other more correlated (possibly even insulating) orbitals. It is found that the slopes of the curves for the upper orbital (the most correlated and closest to half-filling, that can be viewed as mimicking the t2g orbitals in the realistic calculations) scale exactly (see inset in Fig. 1d) with the bare-crystal field splitting (it is worth recalling here that the crystal field renormalized by the interactions changes with the filling, instead). 
 
The physics described in this simplified model is quite similar to the one we have investigated in the ab-initio calculations for iron superconductors.
Indeed this scaling seems to apply to the $t_{2g}$ orbitals in the three compounds studied in Ref. \cite{demedici_OSM_FeSC}: while for the studied iron pnictides (LaFeAsO and BaFe$_2$As$_2$) the $xy$ orbital has a steeper slope than the $xz/yz$ (which have in turn a slope  steeper than the eg orbitals), in FeSe there is an inversion, and the curve for the $xy$ is less steep than for the $xz/yz$.
Now, this seems to reflect the bare crystal fields here too, since as reported in table \ref{fig:Crystal_fields} the bare energy for the $xy$ is above the one for the $xz/yz$ in LaFeAsO and BaFe$_2$As$_2$, while it is below in the chalcogenides. 

Thus it seems that albeit the renormalized crystal-field makes the xy closer to half-filling and hence more correlated than the xz/yz, the footprint of the values of the crystal-field in the non-interacting system remains in the slopes of the $Z_\alpha(n_\alpha)$. Further work is however needed in order to clarify if this is always true, and what is the mechanism behind.

%\input{referenc}
 %\bibliographystyle{spphys}
 %\bibliography{FeSC,publdm,bibldm}

\end{document}